\documentclass[useAMS,usenatbib]{mn2e}

\usepackage[latin1]{inputenc}
\usepackage[english]{babel}
\usepackage{epsfig}
\usepackage{amssymb}
\usepackage{times}
\usepackage{amsmath}
\usepackage{graphicx}
\usepackage{epstopdf}
\usepackage{mathrsfs}
\usepackage{tabularx}
\usepackage{rotating}
\usepackage{longtable,lscape}
\usepackage{natbib}
\newcommand\ion[2]{{#1}\,{\sc #2}} 

\title[Radio quasars at $z \sim 4$]{Neural-network selection of high-redshift radio quasars, and the luminosity function at $z\sim 4$}

\author[D. Tuccillo et al.]{D. Tuccillo,$^1$\thanks{E-mail: tuccillo@ifca.unican.es} 
J. I. Gonz\'alez-Serrano,$^1$ C. R. Benn. $^2$ \\
$^1$Instituto de Fisica de Cantabria (CSIC - Universidad de Cantabria), E-39005, Av. de los Castros s/n Santander, Spain\\
$^2$Isaac Newton Group, Apartado 321, E-38700, Santa Cruz de la Palma, Spain 
}

\begin{document}

\date{accepted for publication in MNRAS on 3 March 2015}

\pagerange{\pageref{firstpage}--\pageref{lastpage}} \pubyear{2011}   

\maketitle

\label{firstpage}

\begin{abstract}

We obtain a sample of 87 radio-loud QSOs in the redshift range $3.6
\le z \le 4.4$ by cross-correlating sources in the FIRST radio survey ($S_{1.4
  GHz} > 1$ mJy) with star-like objects having $r <20.2$ in SDSS Data
Release 7. Of these 87 QSOs, 80 are spectroscopically classified in
previous work (mainly SDSS), and form the training set for a
search for additional such sources. We apply our selection to 2,916
FIRST-DR7 pairs and find 15 likely candidates. Seven of these are
confirmed as high-redshift quasars, bringing the total to 87. The
candidates were selected using a neural-network, which yields 
97\% completeness (fraction of actual high-z QSOs selected as such) and an
efficiency (fraction of candidates which are high-z QSOs)
in the range of 47 to 60\%. We use this sample to estimate the binned
optical luminosity function of radio-loud QSOs at $z\sim 4$, and also
the LF of the {\em total} QSO population and its comoving density.
Our results suggest that the radio-loud fraction (RLF) at high z is
similar to that at low-z and that other authors may be
underestimating the fraction at high-z. 
Finally, we determine the slope of the optical luminosity function and obtain 
results consistent with previous studies of radio-loud QSOs and of the whole
population of QSOs.  The evolution of the luminosity function with redshift 
was for many years interpreted as 
a flattening of the bright end slope, but has recently been re-interpreted as
strong evolution of the break luminosity for high-z QSOs, and our results,
for the radio-loud population, are consistent with this.
\end{abstract}


\section{Introduction}

Since their discovery in the 1960s (\citealt{Hazard1963}; \citealt{Schmidt1963}), quasi-stellar objects (hereafter QSOs) have played a key role in extragalactic research, in particular in connection with studies of super-massive black holes (hereafter SMBHs), galaxy evolution, the intergalactic medium, large-scale structure and cosmology.
 
Quasar candidates are mainly identified from their optical colours in
large sky surveys such as the 2dF survey \citep{Boyle:2000hc} and the
Sloan Digital Sky Survey (SDSS, \citealt{York:2000qo};
\citealt{Schneider:2010lh}). Current data from SDSS provide us with
photometric measurements for $\approx 5 \times 10^8$ galaxies,
quasars and stars. The survey also provides spectra for 
nearly two million of these objects.  This has dramatically increased the
number of known QSOs, since the first
edition of the SDSS quasar catalogue \citep{Schneider2002}.  The 5th
edition, which is based upon the SDSS Seventh Data Release(DR7),
includes a remarkable 105,783 spectroscopically-confirmed QSOs
\citep{Schneider:2010lh}.  Most of the SDSS QSO candidates were selected as
spectroscopic targets on the basis of their non-stellar colours in
$ugriz$ bands and by matching unresolved sources to the FIRST radio
survey (Faint Images of the Radio Sky at Twenty-Centimeters,
\citealt{Becker:1995dz}).


The SDSS QSO-selection algorithm was presented by
\citet{Richards:2002tg}, and according to them is sensitive to QSOs at
redshifts $z <\sim 5.8$.  Completeness (fraction of QSOs selected
as such) and efficiency (number of actual QSOs amongst the candidates, 
divided by the total number of
candidates) of the selection are a complex function of apparent
magnitude $i$ and redshift. Although QSOs of type 2 and certain QSOs
of type 1 are missed, the overall estimated completeness is high
(\citealt{Vanden-Berk:2005kx}), above 90\% for $16.0 \le i \le
19.0$ (\citealt{Richards:2006ye}). At higher redshift, both
completeness and efficiency drop, with an overall completeness of $
\sim 80 \%$ for $3 \le z \le 5.3$ and efficiency $ \sim 55\%$ for QSOs
with $z > 3$.

Determination of the QSO Luminosity Function (QLF) is important for
the study of active galactic nuclei (hereafter AGN) and it requires
QSO samples with a statistically-significant number of sources and
with accurately-known completeness. In particular an accurate
knowledge of the QLF at different epochs allows important constraints
to be placed on the evolution of the mass function of SMBHs, on their
growth and on the lifetime of the QSO phase (\citealt{Volonteri2003};
\citealt{Netzer2007} ; \citealt{Shankar2010} ; \citealt{Shen2012}).
The QLF also provides important information about the impact of QSO
activity on the formation and evolution of the host galaxies
(\citealt{Cattaneo2009}; \citealt{Fabian2012}). In addition it allows
constraint of the contribution of AGN to the X-ray background
(\citealt{Ueda2003}, \citealt{Hickox2006}), the ultraviolet ionizing
radiation (\citealt{Samantaray2000}, \citealt{Giallongo2012}) and the
infrared background \citep{Dole2006}.


Radio-loud QSOs (hereafter RLQs) account for $\sim 8- 13$\% of QSOs
(\citealt{Ivezi2002}; \citealt{Jiang:2007mi};
\citealt{Balokovi2012}). The exact radio-loud fraction (RLF) is still
unknown and it also depends on the definition of
radio-loudness. Usually the latter is based on the radio luminosity $P$
emitted by the source (e.g. \cite{Gregg1996}, define an object to be
radio-loud if $log \, P_{1.4, \rm GHz} (W/Hz) > 25.5$), or on $R$, the
ratio between monochromatic radio and optical luminosities
(\citealt{Stocke1992}). Some authors see no evidence for significant
change of RLF with either redshift or luminosity (e.g.,
\citealt{Goldschmidt1999}; \citealt{Stern2000};
\citealt{Cirasuolo:2003qf}; \citealt{Vigotti:2003vv}), while others
find that the RLF changes with both parameters
(e.g.,\citealt{Miller1990}; \citealt{Visnovsky1992};
\citealt{Schneider1992}; \citealt{Jiang:2007mi};
\citealt{Balokovi2012}). In particular \cite{Jiang:2007mi}, using a
sample of 30,000 optically selected quasars from the SDSS quasar
catalogue obtained from Data Release 3 \citep{Schneider2005}, find
that the RLF of quasars decreases strongly with increasing redshift
and decreasing luminosity.

Although only a small fraction of QSOs are radio-loud and, while they may not be fully representative of the entire population, they can be used to check the completeness of samples selected mainly on the basis of optical data, such as the SDSS QSO catalogues.  For example, selection at radio wavelengths greatly reduces incompleteness due to the effects of dust obscuration, reddening and/or the presence of broad absorption lines (\citealt{Carballo:2008tw}, hereafter C08, and \citealt{McGreer:2009nx}). Selection in the radio is also more {\em efficient}, since it reduces stellar contamination.

In addition, increasing the number of known radio-loud QSOs is important for understanding the origin of the radio phenomenon itself and for clarifying the connection between radio and optical activity in QSOs.  Indeed, whether RLQ and radio-quiet QSOs are two physically distinct populations (the so-called radio-loudness dichotomy) or are part of a continuous sequence is still a matter of debate (see \citealt{Jiang:2007mi}, \citealt{Balokovi2012}, \citealt{kratzer2014}).  As pointed out by several authors (e.g. \citealt{Cirasuolo:2003qf}, \citealt{Balokovi2012}), it would be helpful in the context of this debate, to have more-homogeneous samples of QSOs, a larger number of QSOs
with radio data available, and a reduction of selection biases in flux-limited samples.

In this work we present the results of a selection of RL QSOs with $3.6 \le z \le 4.4$.  We start by cross-matching FIRST radio sources (from the 2003 April 11 release of the catalogue), with star-like objects in the SDSS DR7 photometric catalogue (Section 2). To select candidate high-redshift QSOs (Section 3), we use a simple but reliable neural-network (NN) algorithm already tested in previous work (\citealt{Carballo:2006uq}, \citealt{Carballo:2008tw}).  Spectra of the resulting 15 candidates were obtained with the 2.5m Nordic Optical Telescope (NOT) on La Palma (Spain) and these are presented in Section 4, together with checks of the completeness of our selection and comparisons of the completeness and efficiency with other work. In Section 5 we discuss the K-correction, and  we present our sample of 87 $3.6 \le z \le 4.4$ radio-loud QSOs. The various sources of incompleteness of this sample are discussed in Section 6. In Section 7, we calculate the optical  luminosity function for $z \sim 3.8$ and $z \sim 4.2$. In Section 8 we derive the space density of radio-loud QSOs, and  the space density of all QSOs, and discuss the luminosity function. Our conclusions are summarized in Section 9.

All optical magnitudes are on the AB system. We use a $\Lambda$CDM cosmology with $\Omega_{\lambda} =0.7$, $\Omega_m =0.3$, and $H_0 =70 \, km \, s^{-1} \, Mpc^{-1}$.

\section{Data} 
\subsection {Surveys used}

In this work we use the FIRST radio survey and the SDSS DR7 optical survey to obtain a radio-optical sample of QSO candidates in the redshift range $3.6 \le z \le 4.4$ .

The FIRST survey was originally designed to produce the radio equivalent of the optical Palomar Observatory Sky Survey, using the NRAO Very Large Array (VLA) in its B-configuration at 1.4 GHz.  
Subsequently the survey area was chosen to make it ideal for comparison with the Sloan Digital Sky Survey (SDSS).  The survey produces images with $1\farcs8$ pixels, a resolution of $5''$, a typical rms of 0.15 mJy and a flux-density limit of 1 mJy. The positional accuracy at the survey flux limit is $\sim 1''$.
We  used the 2003 April 11 version of the FIRST catalogue containing  811,117 sources covering a total area of 9,033 deg$^2$ (8,422 deg$^2$ in the Northern Galactic Cap and 611 deg$^2$ in the Southern Galactic Cap).  

The SDSS DR7 \citep{Abazajian:2009ff} covers a total imaging area of 11,663 deg$^2$ (7,646 deg$^2$ in the Northern Galactic Cap). A total of 357 million distinct objects are included in the imaging catalogue, of which approximately 1.6 million are also included in the spectroscopic catalogue.

The survey reaches magnitude limits (95\% detection repeatability for point sources) in photometric bands $u , g, r, i$ and $z$ of $22.0, 22.2, 22.2, 21.3$ and $20.5$ respectively. Absolute astrometric errors are $<0\farcs1$. In this paper, we consider only the images flagged by SDSS as 'Primary'. These are unique detections, i.e. they do not include duplicate detections from the overlap between survey stripes. Each such object is associated with a run and a field which is the primary source of imaging data at this position.

To determine the overlap area of the two surveys we first determined the area of the FIRST survey, which has an irregular boundary,  by constructing a Delaunay triangulation using the source coordinates. We used code developed by \cite{Bernal1988} which provides the coordinates of the vertices of the unique set of triangles over the FIRST area. We then computed the area covered by FIRST by adding up the areas of the individual triangles. The resulting area is 9,032.27 square degrees.
Finally, for each FIRST source, we queried the SDSS database to see if the position of the sources was included in the survey. The result of the query was that 89.38\% of FIRST sources fall in the SDSS-DR7 imaging area, implying an overlap of 8,073.04 square degrees.

\subsection{Pre - selection criteria}
We matched each FIRST source, not flagged as possible sidelobe or nearby bright source ($\sim 3.6\%$ of the sources in the catalog have this warning flag), with the closest optical object in the 'PhotoPrimary' view of the SDSS DR7 catalogue within a $1\farcs5$  radius.
This radius is the same as used by C08, and is a compromise between completeness and efficiency. The adopted value is lower than the $2\farcs0$  radius used by SDSS in their algorithm for QSO selection. However, the excellent astrometry of FIRST and SDSS means that the peak in the distribution of optical/radio offsets occurs at about $0\farcs2$ (\citealt{Schneider:2010lh}, Fig. 6), supporting our adoption of a $1\farcs5$ radius.
From this match we obtained a starting sample of 222,517 sources.
In this sample there is no selection by radio flux density or radio morphology other than the requirement that the radio source have at least a weak core component.  The FIRST catalogue itself introduces several minor selection effects: the FIRST sensitivity limit is somewhat non-uniform over the sky, with small variations due to the observing strategy and large variations due to decreasing sensitivity in the vicinity of bright sources. However, the fraction of the survey area affected by sensitivity variations is small, less than 15\% (\citealt{Becker:1995dz}). Another effect is that the FIRST survey limit of $\geqslant 1.0$ mJy refers to the peak flux density of sources rather than to the integrated flux density; consequently, extended sources with total fluxes greater than 1 mJy may not appear in the catalogue because their peaks fall below the detection threshold.

From these 222,517 matches, we first selected the 13,956 star-like objects with $15.0 \le r \le 20.2$, where r refers to SDSS psfMag\_r, corrected for Galactic extinction according to \cite{Schlegel:1998bh}.

We then filtered the sample on the basis of several SDSS quality-control parameters used by others (e.g. (\citealt{Richards:2002tg})) when selecting QSO targets for spectroscopy. Specifically, we rejected all objects with magnitude errors $> 0.2$ mag in all five bands, and any for which the SDSS 'fatal' error flags 'BRIGHT', 'SATURATED', 'EDGE' or 'BLENDED'  were set, indicating unreliable photometry. This left 13,287 objects.

Finally, in contrast with \cite{Richards:2002tg}, we rejected all
objects with the 'CHILD' flag set (another 4,148 sources), indicating
objects obtained by de-blending an image flagged 'BLENDED'.  This
criterion ensures that only one
optical object is associated with each radio source
(\citealt{Carballo:2006uq}), and we adopt it for consistency with C08.
In this way we avoid introducing
differences in the pre-selection that may change the final efficiency
of the neural-network algorithm.  This is the main source of
incompleteness in our sample, as will be discussed in Section 8.

This pre-selection process, 
summarised in Table \ref{preselection}, left us with 9,139 star-like objects coinciding with FIRST radio sources.

\begin{table}
\caption{Steps in the pre-selection process}
\label{preselection}
\begin{tabular}{@{} lr }
\hline
Pre-selection criterion  & Selected sources  \\ 
\hline
Starting FIRST-SDSSS sample ($\le 1''.5$) & 222,517  \\
15 $\le r \le$ 20.2 & 74,853   \\
Point-like    & 13,956  \\
Without mag error $>0.2$ in all bands & 13,934   \\
Exclude sources with 'fatal' error flags & 13,287  \\
Exclude sources with 'CHILD' flag & 9,139  \\
\hline
\end{tabular}
\end{table}

\section{Selection of QSO candidates}
\subsection{Neural Network algorithm}

The machine-learning technique used in this work to select our list of QSO candidates is described in earlier papers (C06, C08), and here we give only a brief summary. 

We used a supervised Artificial Neural Network (NN)  algorithm of feed-forward type, suitable for solving classification problems, and programmed using the \emph{Matlab Neural Network Toolbox software $^\circledR$}.  A supervised NN is trained with samples of known classification, in order to learn how to distinguish between the classes. Only after the training has been carried out, can the trained NN  be used to classify a new problem sample. In our case the classification task was formulated as a binary problem, the two classes being: (a) the target class, i.e. QSOs in the redshift range $3.6 \le z \le 4.4$, and (b) the 'non-target' class, i.e. all other types of object.


In a feed-forword NN, 
each input variable corresponds to a node in the so-called `input layer', 
Each of the input nodes has a
weighted connection to every node in the next layer, called 'the
hidden layer'. A node in the hidden layer forms a weighted sum of its
inputs, and then passes the information to a second hidden layer that
performs a similar processing. The number of hidden layers, like the
other parameters of the NN, has to be optimised for the problem
that the NN aims to solve. The weighted information passes through the
layers of the NN to the last 'output' layer, which performs a
simple sum of its inputs, giving the output.

 In particular our NN was composed of the input layer, just one hidden layer, and an output layer $y$.  The output $y$ for the $i$th object, with values in the range $(0,1)$, is given by the non-linear function:

\begin{equation}
y^i=\frac{1}{1+e^{-a^i}}
\end{equation}
with  $a^i = w_0 + \sum_{j=1}^d w_j x_j^i$, where ($x_1$,$x_2$,....,$x_d$)$^i$ are the input variables for object i. $w_0$ and ($w_1$, $w_2$, ...,$w_d$), called bias and weights respectively, are the parameters fitted during the training. This NN model is known as \emph{logistic linear discriminant}. The adopted error function was the variance of the outputs:

\begin{equation}
e=\frac{1}{m} \sum_{i=1}^m (y^i-T^i)^2
\end{equation}

where $m$ is the number of objects used for the training and $T$ is the target value, set to 1 for the class of high-redshift QSOs and 0 for the remaining sources, during the training. The optimal parameters for the net, i.e. those minimising the error, were obtained using the Levenberg-Marquardt algorithm. This is a simple but robust function for optimisation and it appears to be the fastest for the training of moderate-sized NNs (\citealt{Hagan1994}).

The set of input variables adopted for the NN is the best set obtained
in C08, i.e. a combination of optical magnitudes and colours
($r$, $u-g$, $g-r$, $r-i$, $i-z$) and 
radio-optical separation. The input variables were
pre-processed, normalising their values to the range ($-1$, $1$). No
outliers required trimming.  For this step the whole pre-selected
sample was used, regardless of whether the source was
spectroscopically classified and thus suitable as a training object, or
not. In fact, the input variables of the new objects (i.e. the problem
objects) presented to the trained net are expected to be normalised in
the same way as the ones used in the training process.

 \subsection{Spectroscopic classification of the pre-selected sample}

Of the sample of 9,139 sources passing initial selection (Table 1), 6,091 have spectra in the SpecObj view of SDSS-DR7. 5,348 of the latter are included in the 5th edition of the SDSS Quasar Catalogue (DR7 QSO Catalogue; \citealt{Schneider:2010lh}), which is also based on SDSS-DR7, but uses more stringent criteria for the classification of the objects as QSOs, in order to exclude dubious cases. 71 of the QSOs in this catalogue have redshifts in the range  $3.6 \le z \le 4.4$. 

The remaining 743 sources with spectra in DR7-SpecObj but not included
in the DR7 QSO Catalogue are classified by SDSS as
stars, galaxies, quasars and sources of
'unknown' type.  
For all of these sources, a search was made in NED (the NASA/IPAC
Extragalactic Database), and none of them was classified there as a $z
\ge 3.6$ QSO. We also visually inspected the DR7-SpecObj spectra of
these sources to check if any of them could be a QSO in the
redshift range of interest here, but none of the objects, which
include those with 'unknown' spectra, had spectral features consistent
with a high-z QSO.

3,048 of the 9139 sources in the sample lack spectroscopic
classification in DR7-SpecObj. However, the DR7 QSO catalogue was compiled by inspecting
all the SDSS spectra, not just those of the quasar candidates, and 
identified 115 of these as quasars (one of them being a high-z QSO in
the redshift range of interest for our work) despite them not being
automatically identified as such. Another 17 sources were identified
as QSOs by C08, comprising 8 high-z QSOs and 9 QSOs with redshift
below 3.6. The remaining 2,916 sources were checked in NED, and none
of them had been spectroscopically classified as of March 2012.

A total of 6,223 (= 6091 + 115 + 17) sources 
thus have a reliable spectroscopic classification.
Of these, 80 (= 71 + 1 + 8) are QSOs in the
redshift range $3.6 \le z \le 4.4$. These sources form the training
sample, i.e. the sample used to train the neural network to
distinguish high-z QSOs in our redshift range of interest from other
objects. Quasars with $3.6 \le z \le 4.4$ play the role of the target
class.The non-target sources include stars, galaxies, QSOs with other
redshifts and objects with spectra classed as 'unknown' but lacking
the features expected for our target sources.

The 2916 sources without available spectra from SDSS-DR7 or from the literature (as of March 2012), 
form the sample from which new QSO candidates in our redshift range of interest are selected using the trained NN.

\subsection{Training and testing of the NN}

Having defined the training sample, with  6223 spectroscopically classified sources, of which 80 are high-redshift QSOs, we are ready to train the NN and to test its performance as a classifier.

The classification algorithm fitted by the NN provides for each source
an output $0 \le y \le 1$. The extreme values of 1 and 0 correspond
respectively to sources with input variables more similar or less
similar to those of the high-z QSO class. Objects with measured $y$
greater than some threshold value $y_c$ are candidate high-z QSOs.

The performance of the trained NN can be expressed in terms of two
basic parameters: efficiency (or reliability) and completeness. In our case
the efficiency is the fraction of candidate high-z QSOs
selected by the NN which are true high-z QSOs. The completeness is the
fraction of true high-z QSOs with $y \ge y_c$ , i.e., the fraction of
true high-z QSOs selected as such by the NN.

\begin{figure}
\centering
\includegraphics[width=0.5 \textwidth]{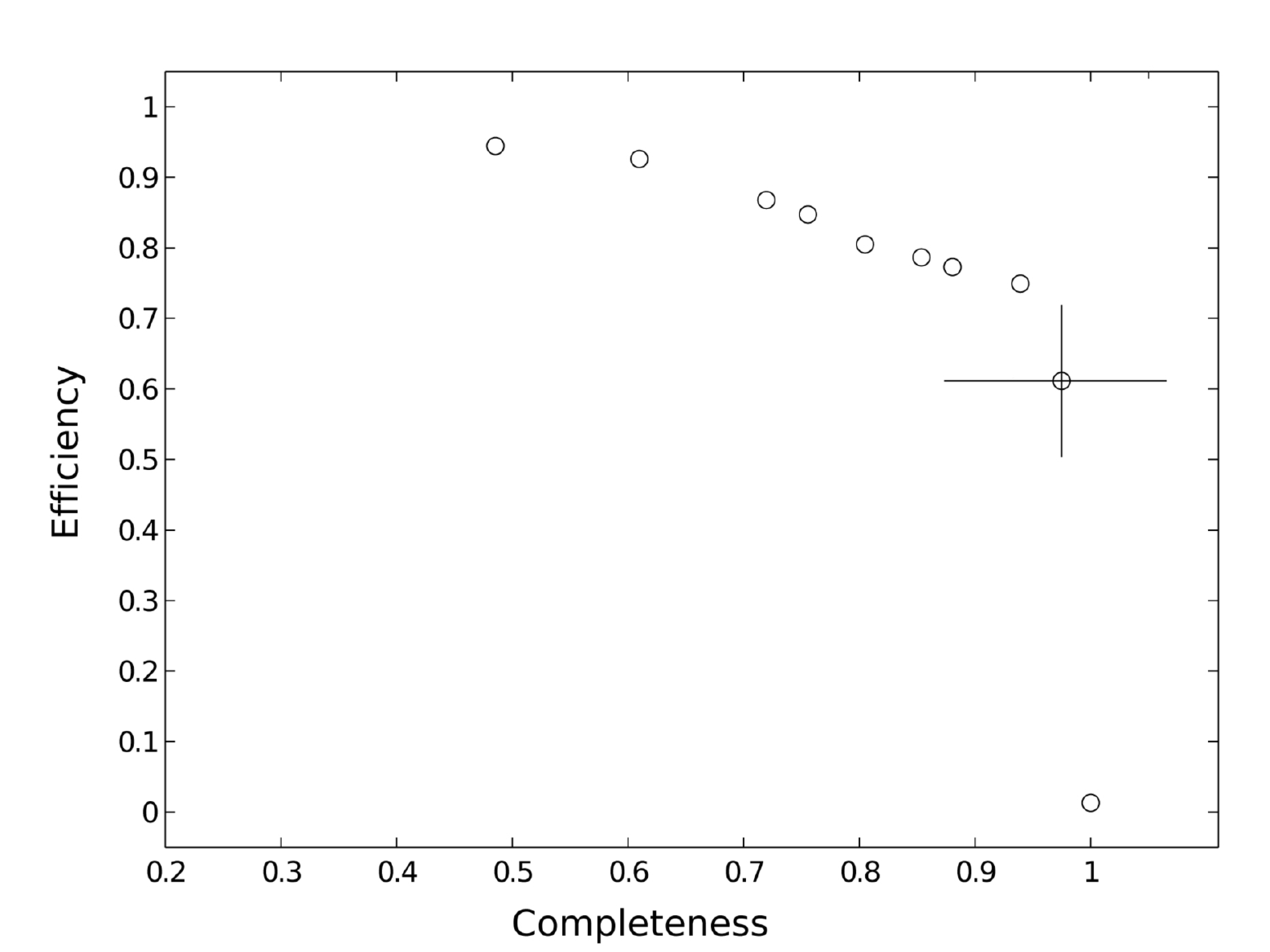}
\caption{Efficiency  versus completeness,  measured for the test sample. Each symbol corresponds to a given value of the threshold $y_c$, ranging from 0.9 to 0, in increments of 0.1, from left to right. The symbol with the error bars correspond to the adopted threshold $y_c$ = 0.1.}
\label{eff_comp}
\end{figure}

The performance of the NN is ideally  tested with a sample of objects not used during the training. In our case, since the target sample has only 80 objects, the `leave one out' method was applied, using all but one of the objects for the training, and  the remaining one for the test. In total 6223 NNs were run, each of them providing  the ouput for a test object. These output values, for a sample of 6223 test objects, were used to compute efficiency and completeness as a function of $y_c$, and the results are shown in Figure  \ref{eff_comp}. Since our purpose is to build a sample appropriate for statistical analysis, priority is given to completeness, accepting lower $y$ values at the cost of lower efficiency. Choosing $y_c = 0.1$ our NN classifier has an efficiency of $60\% \pm 9$ and a completeness of $97\% \pm 11 $ (errors assume Poisson statistics).
 
\subsection{High-redshift QSO candidates}

The set of 6223 trained NNs used for the testing was applied to the sample of 2916 sources without spectra, in order to find candidate high-z QSOs. For each source we adopted the median of the 6223 output values, and  we selected as high-redshift QSO candidates those sources with $y_{med} > 0.1$. In this way 15 QSO candidates in the range  $3.6 \le z \le 4.4$ were found (Table 2), out of an original set of 2,916 sources lacking spectroscopic identification.

\section{Checks of the QSO selection}
Below we present spectroscopy of the 15 NN-selected high-z-QSO candidates (Section 4.1) and use SDSS DR9 to check the completeness of our selection (Section 4.2). In addition, we use our new data to assess the efficiency of high-z-QSO selection by C08 (Section 4.3) and to check the completeness of the spectroscopic identification of high-z-QSOs in SDSS DR7 (Section 4.4).

\subsection{Spectroscopy of the 15 NN-selected candidates}

At the time, none of the 15 QSO candidates selected by our NN had
spectra available in the literature. Therefore long-slit spectra of all 15
were obtained with the 2.5-m Nordic Optical Telescope (NOT)
during the night of 25th March 2012, using ALFOSC 
(Andaluc\'\i a Faint Object Spectrograph and Camera) with
grism \#4, which provides a dispersion of 3\AA\ per pixel.  The
spectral coverage was $4000- 9000$ \AA\ and the resolution was
$15$\AA.  The exposure time was about 1000s per source, delivering
signal-to-noise ratios $\gtrsim 8 $ per pixel.  The seeing was typically
better than 1''.3 FWHM. A spectrophotometric standard star was
observed to correct for spectral response. After each target spectrum,
an exposure of an arc-lamp was taken for wavelength calibration.  Data
were reduced using standard IRAF\footnote{IRAF is distributed by the
  National Optical Astronomy Observatories, which is operated by
  Association of Universities for Research in Astronomy, Inc., under
  cooperative agreement with the National Science Foundation.}
routines.

\begin{figure*}
\centering
\includegraphics[width=0.7 \textwidth]{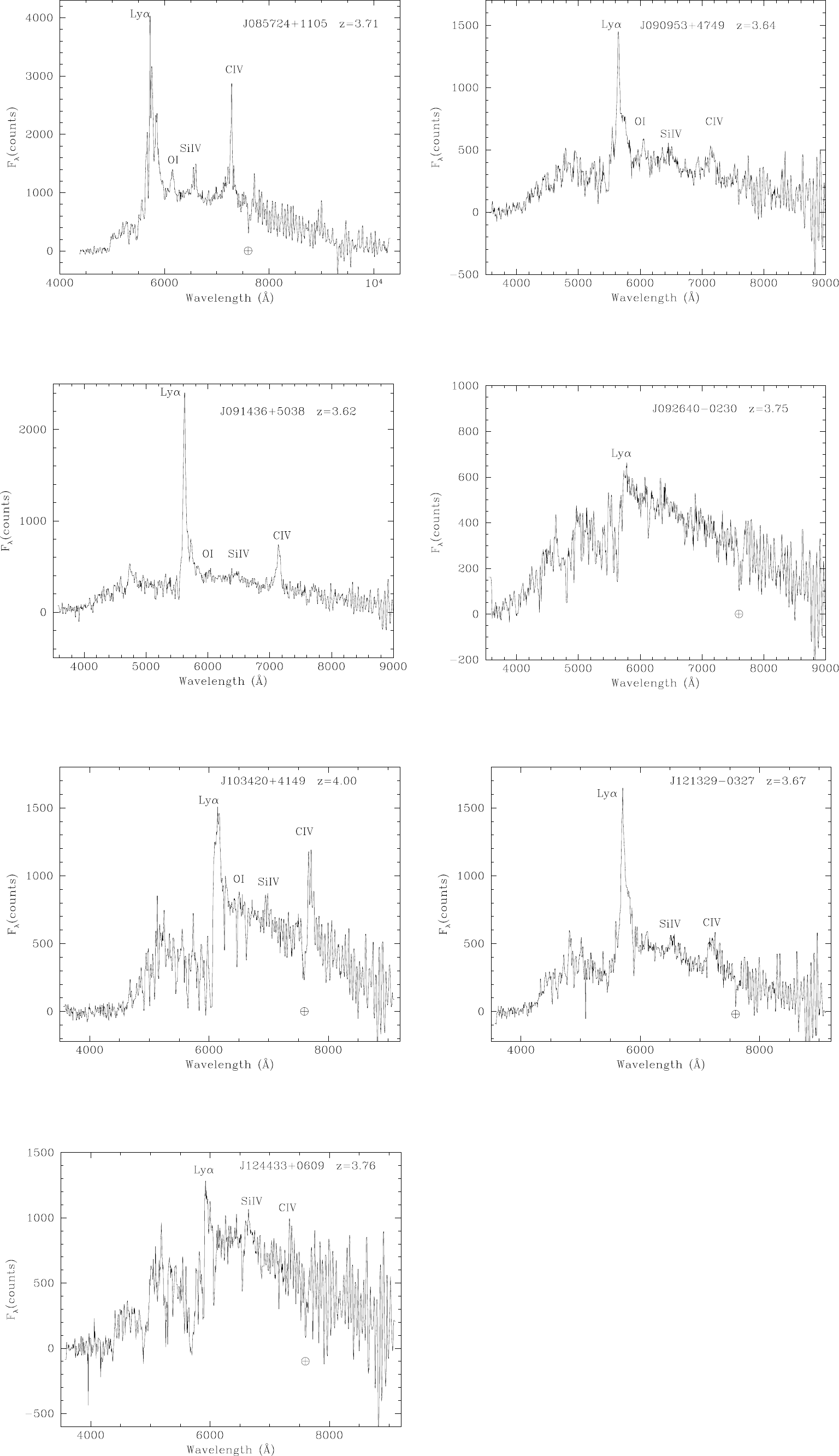}
\caption{NOT/ALFOSC spectra of NN-selected high-z candidates: QSOs with $3.6 \le z \le 4.4$}
\label{mag3e6}
\end{figure*}

\begin{figure*}
\centering
\includegraphics[width=0.7 \textwidth]{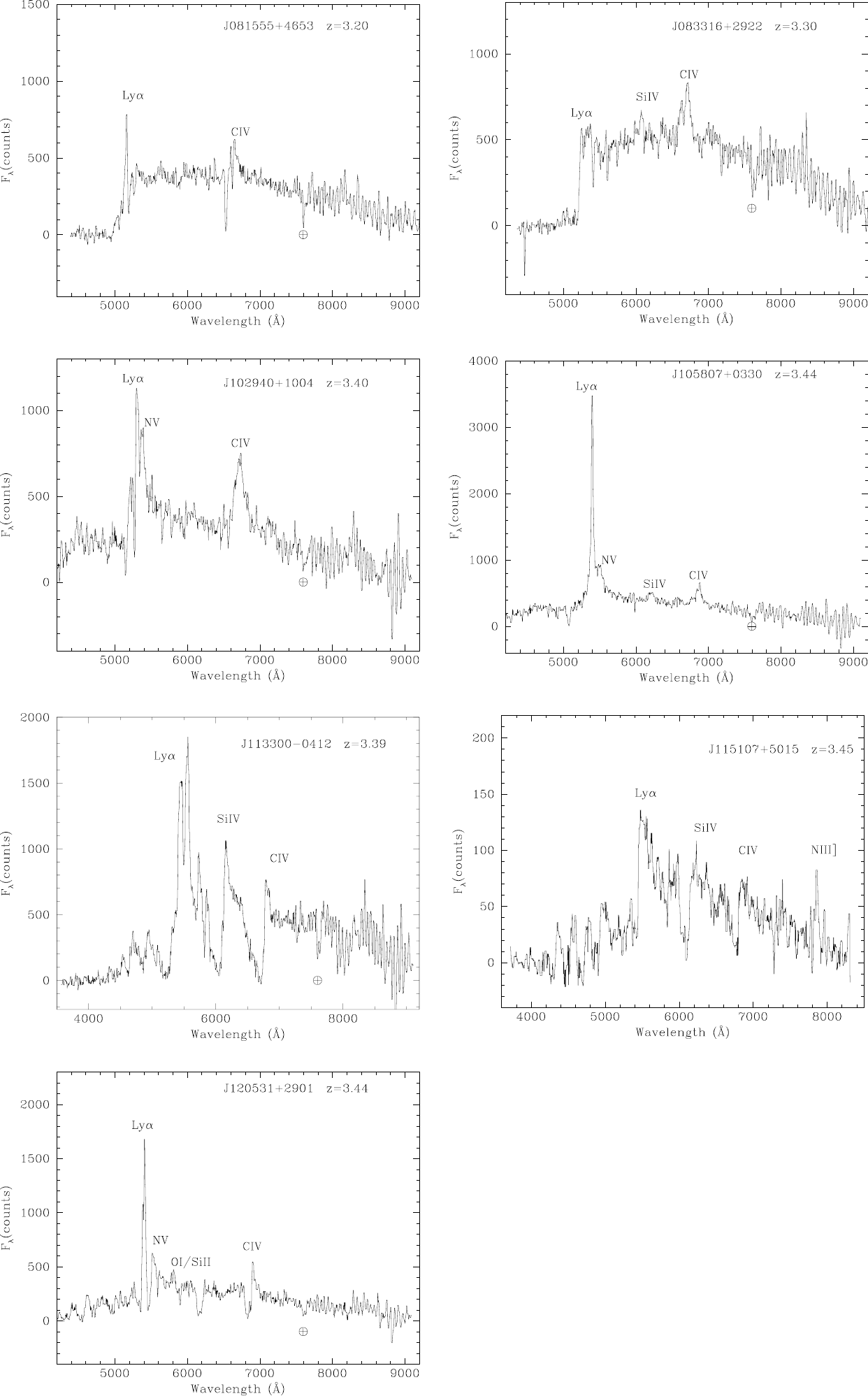}
\caption{NOT/ALFOSC spectra of NN-selected high-z candidates: QSOs with $3.0 \le z < 3.6$}
\label{spectra_others}
\end{figure*}

\begin{figure*}
\centering
\includegraphics[width=0.7 \textwidth]{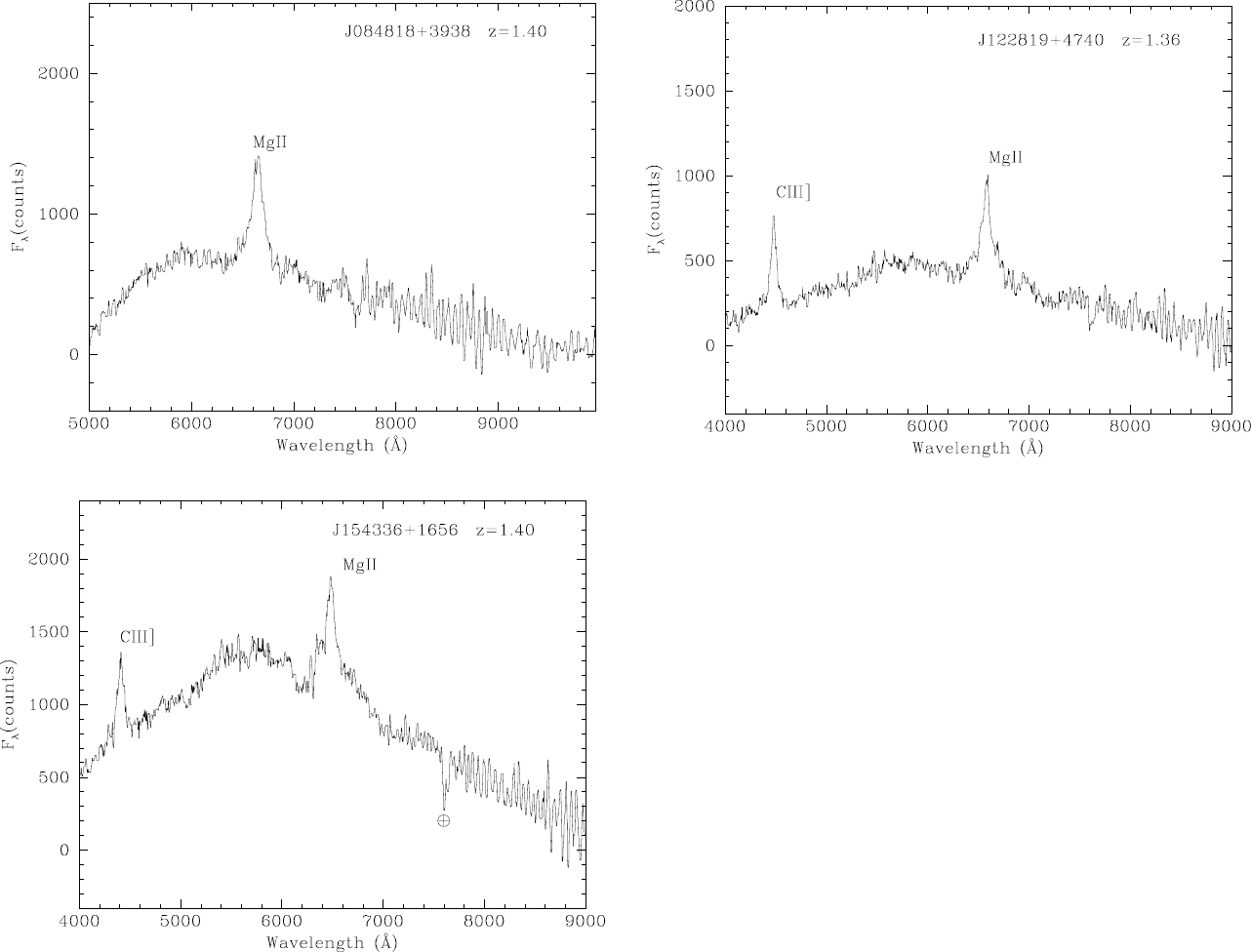}
\caption{NOT/ALFOSC spectra of NN-selected high-z candidates: QSOs with $z < 3$}
\label{spectra_others2}
\end{figure*}

All 15 candidates were confirmed as QSOs, 7 of them
(Fig. \ref{mag3e6}) in the desired range of redshift, 5 in the nearby
range $3.1\le z < 3.6$ (Fig. \ref{spectra_others}), and the remaining
3 at lower redshifts (Fig. \ref{spectra_others2}). For the last 3
quasars at lower redshift, the NN-misclassification is caused most
likely by confusion of  \ion{Mg}{ii} emission with Ly$\alpha$
emission.  Table 2 lists the coordinates, $r$ magnitudes, $y_{med}$
values and redshifts for the 15 candidates. A note in column 6
indicates whether there is a spectrum of the object in the latest SDSS
release DR9 (see next subsection).  We report also the discovery of
five new broad absorption line (BAL) QSOs (see Table 2).

The observed efficiency of this selection is therefore $ \sim 47\%$ (7 $z >$ 3.6 QSOs out of 15 candidates).

\begin{table*}
\caption{Sample of 15 FIRST-SDSS DR7 high-z QSO candidates selected by our NN}
\begin{tabular}{ r r r  r r r  r r r r }
\hline
\multicolumn{1}{c}{RA} &
\multicolumn{1}{c}{DEC} &
\multicolumn{1}{c}{$r_{AB}$} &
\multicolumn{1}{c}{$S_{1.4\rm{GHz}}$} &
\multicolumn{1}{c}{$y_{med}$} &
\multicolumn{1}{c}{redshift} &
\multicolumn{1}{c}{DR9} &
\multicolumn{1}{c}{C08} &
\multicolumn{1}{c}{Notes} \\
\multicolumn{2}{c}{(J2000)} &
\multicolumn{1}{c}{} &
\multicolumn{1}{c}{(mJy)} &
\multicolumn{1}{c}{} &
\multicolumn{1}{c}{} &
\multicolumn{1}{c}{} &
\multicolumn{1}{c}{} &
\multicolumn{1}{c}{} \\
\multicolumn{2}{c}{(1)} &
\multicolumn{1}{c}{(2)} &
\multicolumn{1}{c}{(3)} &
\multicolumn{1}{c}{(4)} &
\multicolumn{1}{c}{(5)} &
\multicolumn{1}{c}{(6)} &
\multicolumn{1}{c}{(7)} &
\multicolumn{1}{c}{(8)}  \\
\hline
08:15:55.02  & +46:53:21.4 & 19.89  & 2.97      & 0.12   &  3.20  &        & yes &  BAL\\
08:33:16.91 & +29:22:28.0  & 20.13  & 12.63    & 0.32   &  3.30  & yes & yes &         \\
08:57:24.33 & +11:05:49.2  & 19.81  & 1.91      & 0.98   &  3.71  &        &       &         \\
09:09:53.85 & +47:49:43.2  & 19.90  & 373.29  & 0.22   &  3.64  &        & yes &         \\
09:14:36.23 & +50:38:48.5  & 20.19  & 47.98    & 0.15   &  3.62  &        & yes &         \\
09:26:40.29 & -02:30:41.5   & 19.82  & 1.9        & 0.12   &  3.76  & yes &        &         \\
10:29:40.93 & +10:04:10.9  & 19.47  & 2.81      & 0.22   &  3.40  &        & yes &         \\
10:34:20.43 & +41:49:37.5  & 20.12  & 2.17      & 0.30   &  4.00  & yes & yes  &         \\
11:33:00.71 & -04:11:58.5   & 19.96  & 9.64      & 0.12   &  3.39  &        &       & BAL \\
11:51:07.42 & +50:15:58.6  & 20.09  & 1.69      & 0.32   &  3.40  &        & yes & BAL \\
12:05:31.73 & +29:01:49.2  & 20.17  & 1.51      & 0.50   &  3.44  &        & yes & BAL \\
12:13:29.43 & -03:27:25.7   & 19.64  & 23.37    & 0.76   &  3.67  &        & yes &         \\
12:28:19.97 & +47:40:30.4  & 19.32  & 2.24      & 0.46   &  1.40  &        & yes &         \\
12:44:43.07 & +06:09:34.6  & 19.78  & 1.29      & 0.21   &  3.76  & yes & yes &         \\
15:43:36.59 & +16:56:21.8  & 18.97  & 10.85    & 0.12   &  1.40  &        &       & LoBAL\\
\hline
\end{tabular}

\medskip
The columns give the following: (1) SDSS J2000 coordinates; (2) SDSS dereddened PSF $r$ magnitude; (3) FIRST peak radio flux density; (4) NN output; (5) QSO redshift determined in this work; (6) indicates if the source has a spectrum in SDSS-DR9; (7) indicates if the source was previously selected as a high-z candidate by C08;  
(8) BAL - broad-absorption-line QSO; LoBAL - low-ionization broad-absorption-line QSO. 

\label{15_candidates}
\end{table*}

\begin{table*}
\caption{Spectra of three high-z QSO candidates from C08, not selected by our NN, but observed to complete the C08 sample}
\centering
\begin{tabular}{r r r r r r r}
\hline
\multicolumn{1}{c}{RA} &
\multicolumn{1}{c}{DEC} &
\multicolumn{1}{c}{$r_{AB}$} &
\multicolumn{1}{c}{$S_{1.4\rm{GHz}}$} &
\multicolumn{1}{c}{redshift} &
\multicolumn{1}{c}{DR9} \\
\multicolumn{2}{c}{(J2000)} &
\multicolumn{1}{c}{} &
\multicolumn{1}{c}{(mJy)} &
\multicolumn{1}{c}{} &
\multicolumn{1}{c}{}  \\
\multicolumn{2}{c}{(1)} &
\multicolumn{1}{c}{(2)} &
\multicolumn{1}{c}{(3)} &
\multicolumn{1}{c}{(4)} &
\multicolumn{1}{c}{(5)} \\
\hline
08:48:18.88 &  +39:38:06.0 & 20.15  & 1.28 & 1.4   &   \\
10:58:07.47 & +03:30:59.6 & 19.91  & 4.18 & 3.44 & yes  \\
12:04:07.84 & +48:45:48.0 & 19.97  & 3.96 & M-star  &  \\
\hline\end{tabular}

\medskip
The columns give the following: (1) SDSS J2000 coordinates; (2) SDSS dereddened PSF $r$ magnitude; (3) FIRST peak radio flux density; (4) QSO redshift or spectral classification;  (5) indicates if the source has a spectrum in SDSS-DR9.

\label{7_more}
\end{table*}

\subsection{Check of completeness from SDSS DR9}

As discussed in earlier sections, our methodology is applied to a
sample of sources detected in both SDSS and FIRST using a match
radius of $1\farcs5$, and it makes use of photometric data in SDSS and
of the radio-optical separation. Our pre-selection and selection
methods are therefore based on the same variables as the SDSS
target-selection algorithms. Therefore we checked if the more recent
SDSS-DR9 spectroscopic catalogue provides spectra of any of the 2,916
sources lacking a spectral classification after SDSS-DR7 was released.

 The 'SpecObj' view of SDSS-DR9 was used for this purpose, giving the following results. 4 of the 15 candidates have spectra in DR9 (see Table \ref{15_candidates}) with redshifts very similar to those reported in this work. Of the remaining 2,901 sources rejected by the NN as high-z QSO candidates, 451 have spectra in DR9 and 4 of them are classified as QSOs with $3.6 \le z \le 4.4$. However, examination of the spectra reveals that all four are actually lower-redshift objects or stars.
These objects are J075757.87+095607.56, J101403.75+451053.27, J112742.74+363429.5, and J222758.13+003705.45.

\begin{figure}
\centering
\includegraphics[width=0.5 \textwidth]{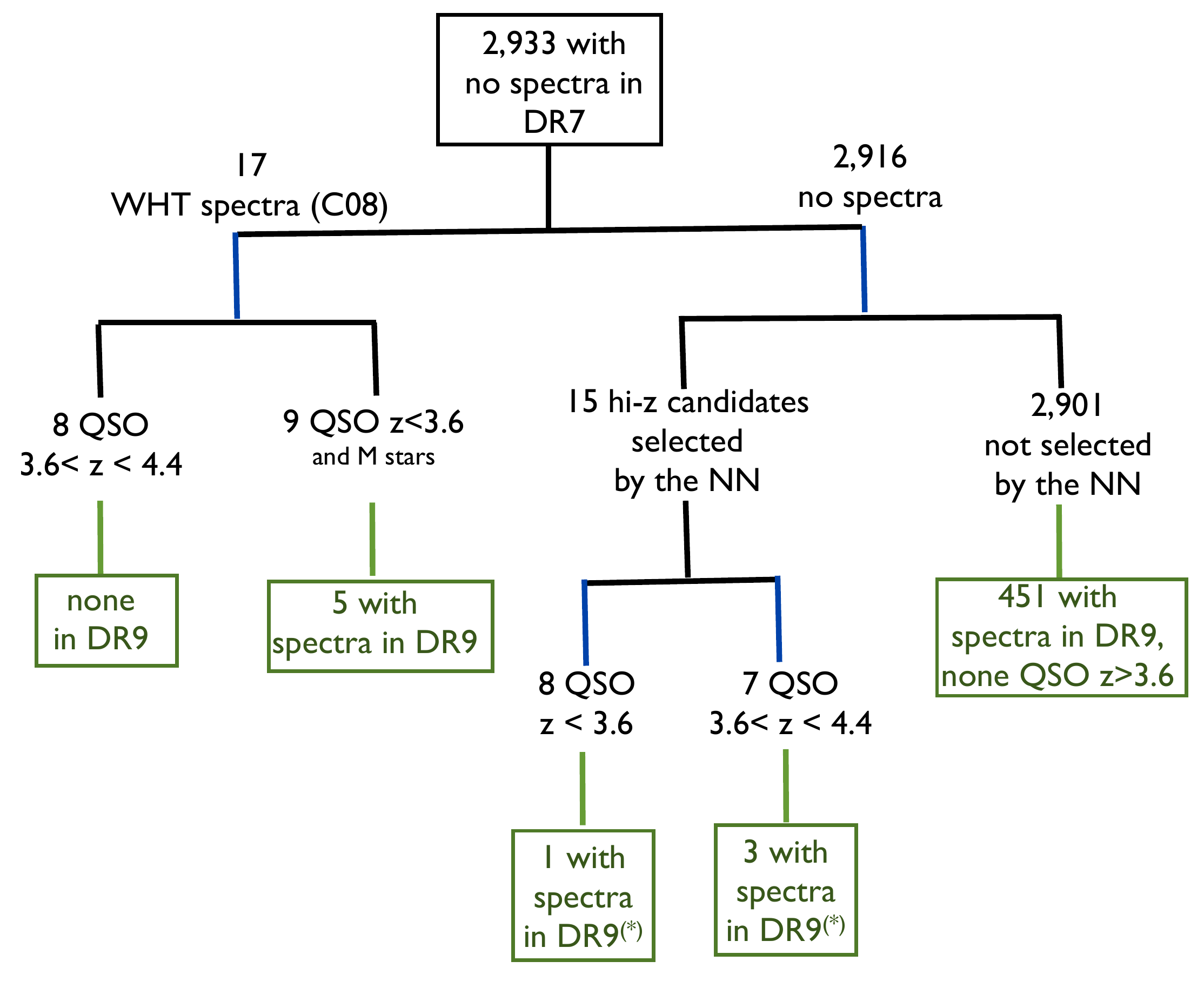}
\caption{
Schematic overview of a check of the efficiency of our selection
via comparison with DR9 spectroscopic information. See Section
4.2.  The 2916 objects (Section 3.2) without spectra in DR7, or in the literature,
were classified by our NN, with 15 being selected by the NN as high-redshift
candidates.
(*) = targets from BOSS (Baryon Oscillations Spectroscopic
Survey).}
\label{schema2}
\end{figure}

The fact that none of these 451 objects are classified in DR9 as
high-z QSOs is 
consistent with the estimated high completeness of our NN algorithm:
$\sim 97 \%$.

\subsection{Assessment of the sample of high-z candidates in C08}

The selection of high-z QSOs in the present paper is based on more recent SDSS data releases than used by C08. However the FIRST-SDSS pairs were obtained in this work in same way as by C08, with similar criteria for the magnitude limits, the maximum radio-optical separation, optical morphology, and photometric quality. Also the classification procedure, aimed at the identification of new high-z QSOs, uses similar NN architecture, input parameter-set, NN training and output parameter.

In C08 we used a training sample comprising 52 QSOs with $z \ge 3.6$, all from the DR5 spectroscopic catalogue, and selected 58 new candidates. In that paper 24 of the candidates were confirmed as $z \ge 3.6$ QSOs (17 from observations obtained for the paper and 7 from the literature or from the DR6 spectroscopic catalogue).  16 sources were classified as other types of object, on the basis of observations by C08, or spectra from the literature or DR6.  18 high-z QSO candidates remained unclassified at C08. 

11 of the 18 C08 candidates overlap with the 15 candidate-QSOs identified in this work.   As result of our observations, these 11 sources were classified as 5 high-z QSOs, 5 QSOs with $3.2 \le z \le 3.5$ and a QSO at $z=1.40$ (see Table \ref{15_candidates}). The remaining seven candidates of C08 consist of three QSOs included in the DR7 QSO Catalogue and in our training sample (SDSS 110946.44+190257.6 with z=3.67, SDSS 123128.22+184714.3 with z=3.33 and SDSS 124323.16+235842.2 with z=3.49) and four sources not selected by our classifier as being high-z QSOs. The spectra of three of these four sources were obtained in our observing programme at the NOT, yielding a QSOs at $z=3.44$ and $z=1.4$ and a late-type star  (see Table \ref{7_more}, and figures \ref{spectra_others} and \ref{spectra_others2} for spectra) . 

The spectroscopic observations of the 58 high-z candidates at C08 are now almost complete (57 out of 58). They yield an efficiency of $52 \pm 9$ per cent (30/58), highlighting the value of simple neural networks for this classification task. This efficiency is in reasonable agreement with the value obtained with the training sample, i.e. with the expectation we had from the objects with known classification, $62 \pm 9$ per cent (C08).  In addition, we note that a large fraction of the contaminants, 15 out of 28, are QSOs with $3.1 \le z < 3.6$, close to the redshift threshold we adopted.

\subsection{Spectroscopic completeness of SDSS for high-z QSOs}

Several studies of the SDSS selection of QSOs (\citealt{Richards:2002tg}, \citealt{Croom:2004ij}, \citealt{Richards:2006ye}, \citealt{McGreer:2009nx}) suggest an overall completeness above 90\%.  The completeness in general decreases with increasing redshift and decreasing brightness, and it is particularly inefficient for $2.2 < z < 3$ where quasar and star colours are very similar. In particular, \cite{McGreer:2009nx} studied
the completeness of quasar selection at redshift $z>3.5$ and magnitude $i<20.2$, and found for the SDSS target algorithm a completeness of $ \approx 86\%$, in good agreement with the 85\% derived in \cite{Richards:2006ye}.

The analysis in this paper and in C08 identifies 15 QSOs with $3.6 \le z \le 4.4$, missed by SDSS-DR7 (two have spectra in SDSS-DR9,  which uses BOSS).  This allows us to estimate the incompleteness of the SDSS-DR7 selection, after allowing for the fact that the spectroscopic area of the SDSS survey is $\sim 95\%$ of the imaging area. 4 of our 15 QSOs $3.6 \le z \le 4.4$ lie outside the SDSS  spectroscopic plates, so the estimated incompleteness in SDSS is 11 QSOs  out of 83 (87 minus 4) in this redshift range, i.e. $\sim 13 \% \pm 4$, in good agreement with  estimates by \cite{McGreer:2009nx} and \cite{Richards:2002tg}.

\begin{figure}
\centering
\includegraphics[width=0.5 \textwidth]{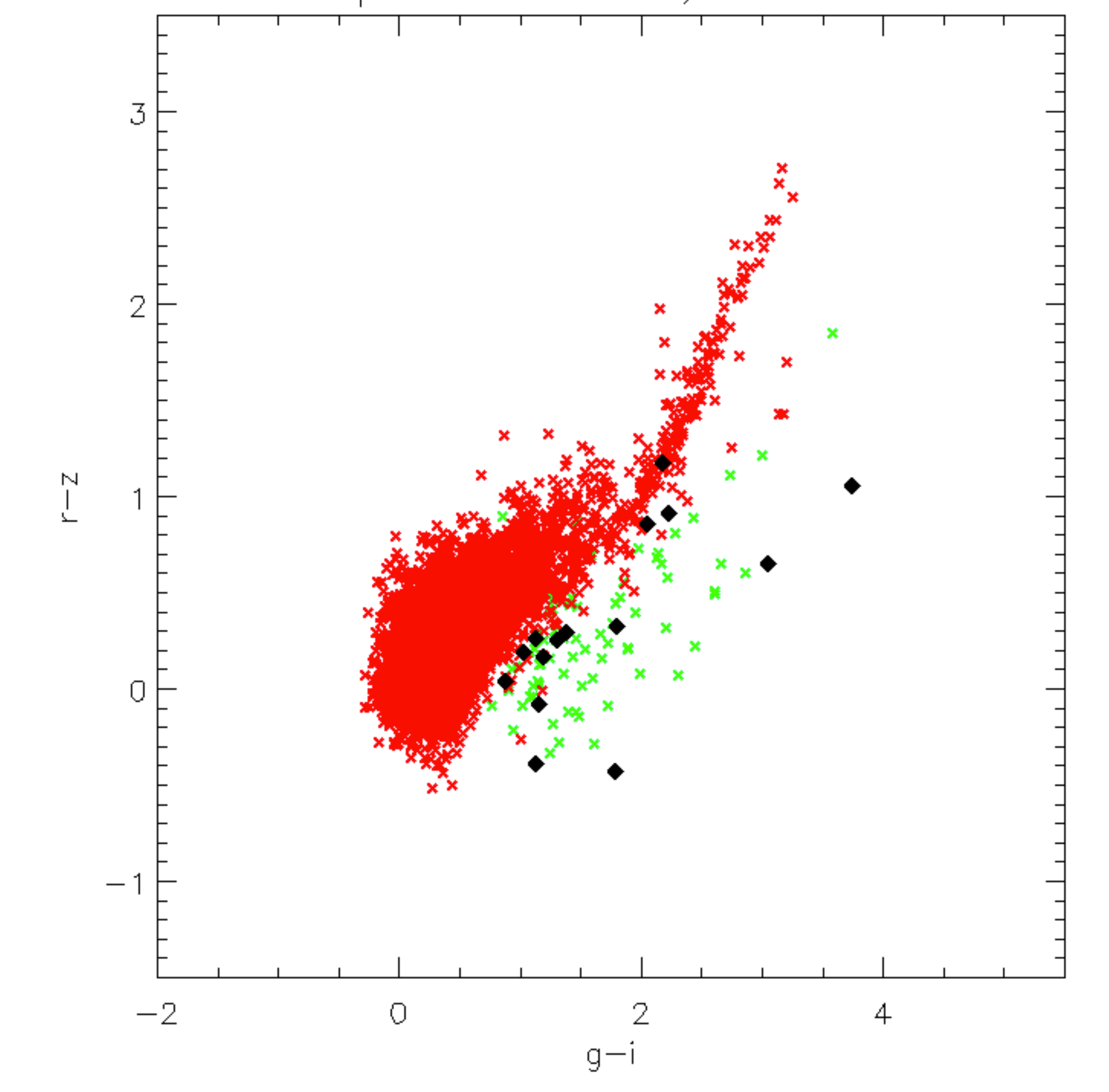}
\caption{Colour-colour diagram of the 6233 sources with spectra in DR7,
with green crosses representing QSOs with $3.6 \le
  z \le 4.4$, and red crosses representing other objects.
The new 15 QSOs with $3.6 \le z
  \le 4.4$
(8 from C08, 7 from this work, all missed by SDSS)
are plotted as black diamonds.}

\label{ColorColor}
\end{figure}

In Fig. \ref{ColorColor} we plot the $g-i,r-z$ colour-colour diagram of the entire training sample together with the 15 $3.6 \le z \le 4.4$
QSOs missed by SDSS and identified by us.  Splitting the sample
between QSOs with $3.6 \le z \le 4.4$ and all other sources, we note
that our QSO-selection method is sensitive even at the boundary of the
two samples, demonstrating the effectiveness of learning-machine
techniques, when compared with simple colour-cut criteria.

\section{Final sample}

In Table \ref{largeTable} we present the final sample of 87 QSOs with
$ 3.6 \le z \le 4.4$ satisfying our selection criteria. This
sample includes 72 QSOs (see Section 3.2) from the 5th quasar catalogue
\citep{Schneider:2010lh} plus 15 QSOs revealed by our neural-network
(8 from C08 and 7 from this work, see Section 4.1).  The magnitide limit for
our sample is $r_{AB}$ = 20.2.  To convert to absolute magnitudes
$M_r$, a $K$-correction (\citealt{Humason:1956oq}) is required.

Following the convention of \cite{Hogg:2002fu}, the $K$-correction between a bandpass $R$ used to observe a source at redshift $z$ and the same bandpass in the rest-frame, is:

\begin{equation}
m_R =  M_R + D_M(z) + K_{R}(z)
\label{conversione}
\end{equation}

\noindent
where $D_M(z)$ is the distance modulus calculated from the luminosity distance $D_L$ as $D_M = 5 \, log_{10} \left ( \frac{D_L}{10 pc} \right)$, $m_R$ is the apparent magnitude, and $M_R$ is the absolute magnitude.

\begin{figure}
\centering
\includegraphics[width=0.5 \textwidth]{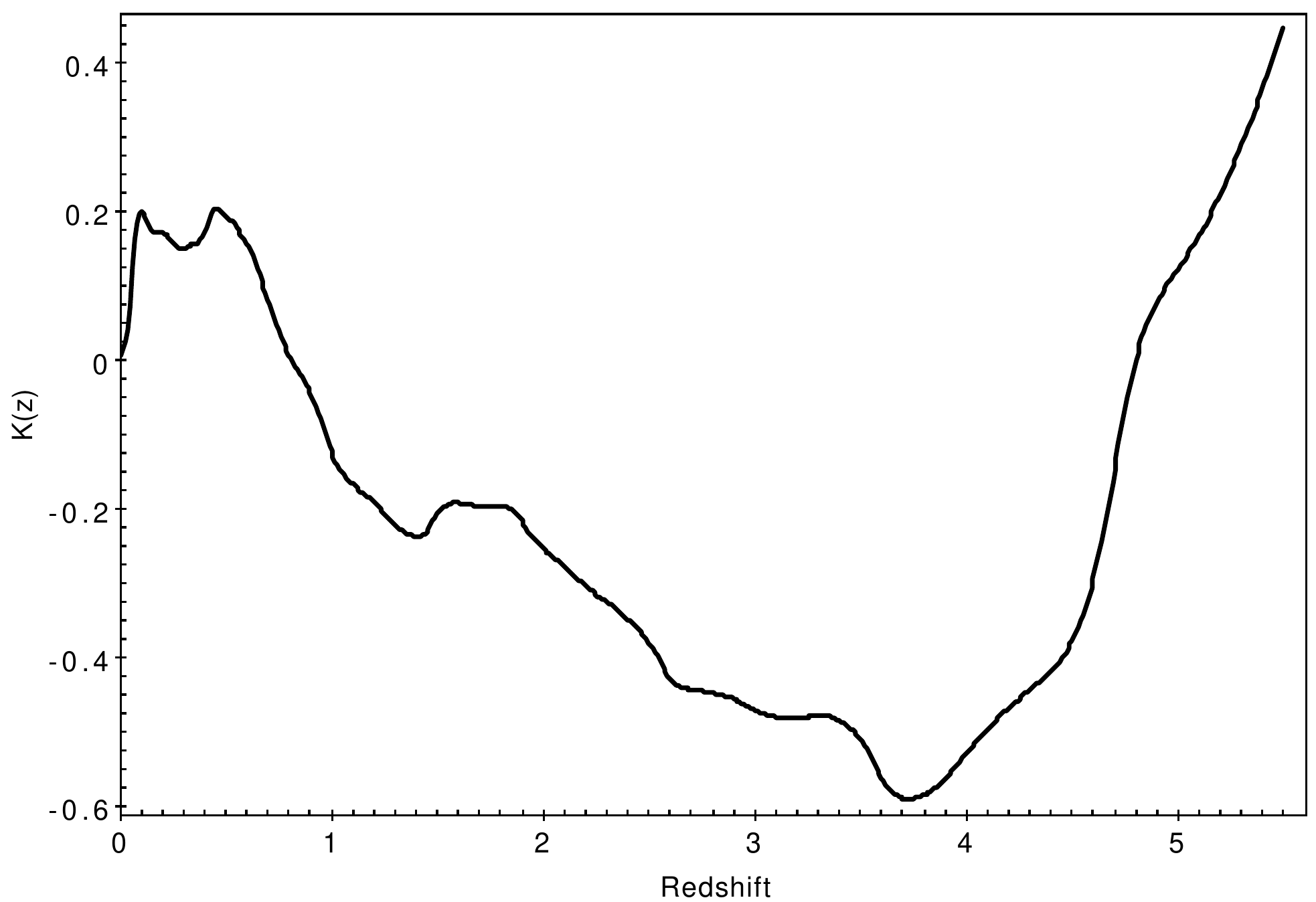}
\caption{Computed $K$-correction in $r$.  A tabulation is available on-line, and a sample
is shown in Table 5.}
\label{Kcorr}
\end{figure}

An accurate  $K$-correction, including the contribution of emission lines, can be computed by convolving a typical QSO spectrum at different redshifts with the filter response (e.g. \citealt{Cristiani:1990fv}, and  \citealt{Wisotzki:2000cr}, the latter based on optical/UV spectra from \citealt{Elvis:1994ys}). Following this approach, we calculated the K-correction by convolving the \cite{Vanden-Berk:2001zr} composite quasar spectrum with the SDSS r-filter:

\begin{equation}
K=   2.5 {\rm log}_{10}\left[(1+z)  \frac{\int_0^\infty F(\lambda) S(\lambda) d \lambda}{\int_0^\infty F((\lambda)/(1+z))S(\lambda) d \lambda} \right]
\end{equation}
 
\noindent

where $F(\lambda)$ is the measured intensity per unit wavelength, and
$S(\lambda)$ is the r-band filter response. The resulting
K-correction is shown in Fig. \ref{Kcorr} and in Table \ref{Kcorr2}.

To convert from  $M_r$ to the commonly used monochromatic absolute AB magnitude at $1450$\AA, $M_{1450}$,  we assume the canonical power-law spectral energy distribution with spectral index $\alpha_{\nu} =-0.5$ and hence:

\begin{equation}
M_{1450} = M_r + 2.5 \alpha_\nu {\rm log}_{10}\left(\frac{1450 \, \AA }{6231 \, \AA}\right) =M_r + 0.791
\label{transf1450}
\end{equation}

\noindent
where $6231$\AA\ is the effective wavelength of the SDSS $r$ filter. We use this spectral index
rather than that derived by \cite{Vanden-Berk:2001zr} ($\alpha_{\nu} =-0.44$) to allow a direct comparison with other authors. We also avoid using the composite spectrum in \cite{Vanden-Berk:2001zr} to convert from 6231\AA\ to 1450\AA\ rest-frame flux density since above H$\beta$ there is a significant contribution from stellar light from  low-redshift quasar hosts, which we effectively eliminate by extrapolating the power-law to red wavelengths.

\begin{table*}
\centering
\caption{Final sample of 87 radio-loud QSOs with $3.6 \le z \le 4.4$ (continued on next page)}
\label{largeTable}
\begin{tabular}{r r r r r c r r c r }
\hline
\multicolumn{1}{c}{NAME} &
\multicolumn{1}{c}{RA} &
\multicolumn{1}{c}{DEC} &
\multicolumn{1}{c}{$r_{AB}$} &
\multicolumn{1}{c}{$\sigma_r$} &
\multicolumn{1}{c}{Redshift} &
\multicolumn{1}{c}{$S_{1.4\rm{GHz}}$} &
\multicolumn{1}{c}{$M_r$} &
\multicolumn{1}{c}{log$_{10}P_{1.4\rm{GHz}}$}  &
\multicolumn{1}{c}{ID} \\
\multicolumn{1}{c}{} &
\multicolumn{2}{c}{(J2000)} &
\multicolumn{1}{c}{} &
\multicolumn{1}{c}{} &
\multicolumn{1}{c}{} &
\multicolumn{1}{c}{(mJy)} &
\multicolumn{1}{c}{} &
\multicolumn{1}{c}{(W/Hz)}  &
\multicolumn{1}{c}{}  \\
\multicolumn{1}{c}{(1)} &
\multicolumn{2}{c}{(2)} &
\multicolumn{1}{c}{(3)} &
\multicolumn{1}{c}{(4)} &
\multicolumn{1}{c}{(5)} &
\multicolumn{1}{c}{(6)} &
\multicolumn{1}{c}{(7)} &
\multicolumn{1}{c}{(8)}  &
\multicolumn{1}{c}{(9)} \\
\hline
J015339.61$-$001104.9 &  01:53:39.61 & $-$00:11:05.0 & 18.83 & 0.022 & 4.19  & 4.82   & -28.59 &  26.42  & SDSS  \\ 
J030025.23+003224.2  &  03:00:25.23 & +00:32:24.2 & 19.68 & 0.025 & 4.18 & 7.75     & -27.74	& 26.62 & SDSS \\
J072518.26+370518.3 &  07:25:18.27 & +37:05:18.4 & 19.60 & 0.020 & 4.33 & 26.72   & -27.94	& 27.19 & WHT \\
J074711.14+273903.3 &  07:47:11.15 & +27:39:03.4 & 18.35 & 0.025 & 4.15 & 1.08     & -29.04	&  25.76 & SDSS \\
J074738.49+133747.3 &  07:47:38.49 & +13:37:47.3 & 19.35 & 0.015 & 4.17 & 7.18     & -28.05	& 26.59 & WHT \\
J075113.04+312038.0 &  07:51:13.05 & +31:20:38.0 & 19.73 & 0.020 & 3.76 & 5.84     & -27.29	& 26.42 & SDSS \\
J075122.35+452334.2 &  07:51:22.36 & +45:23:34.2 & 20.20 & 0.033 & 3.61 & 1.18     & -26.73	& 25.69 & SDSS \\
J080710.74+131739.4 &  08:07:10.74 & +13:17:39.4 & 20.00 & 0.026 & 3.73 & 48.20     & -27.00	& 27.32 & SDSS \\
J081009.95+384757.0 &  08:10:09.95 & +38:47:57.1 & 19.62 & 0.018 & 3.94 & 26.68   & -27.57	& 27.11 & SDSS \\
J082323.32+155206.8 &  08:23:23.32 & +15:52:06.8 & 19.30 & 0.018 & 3.78 & 74.93   & -27.74	& 27.53 & SDSS \\
J083322.50+095941.2 &  08:33:22.50 & +09:59:41.2 & 18.69 & 0.016 & 3.73  & 122.52 & -28.31	& 27.73 & SDSS \\
J083808.46+534809.8 &  08:38:08.46 & +53:48:09.9 & 19.94 & 0.032 & 3.61 & 8.47     & -27.00	& 26.54 & SDSS \\
J083946.22+511202.8 &  08:39:46.22 & +51:12:02.9 & 19.31 & 0.016 & 4.39 & 40.50     & -28.28	& 27.38 & SDSS \\
J084044.19+341101.6 &  08:40:44.19 & +34:11:01.6 & 19.78 & 0.020 & 3.89 & 13.64   & -27.35	& 26.81 & SDSS \\
J085257.12+243103.1 &  08:52:57.12 & +24:31:03.2 & 19.46 & 0.016 & 3.62 & 157.30   & -27.48	& 27.81 & SDSS \\
J085501.81+182437.7 & 08:55:01.82 & +18:24:37.7 & 19.96 & 0.020 & 3.96 & 9.38      & -27.25	 & 26.66 & SDSS \\
J085724.32+110549.1 &  08:57:24.33 & +11:05:49.2 & 19.81 & 0.017 & 3.71 & 1.91     & -27.17	& 25.92 & NOT  \\
J085944.06+212511.1 &  08:59:44.06 & +21:25:11.2 & 18.74 & 0.015 & 3.70 & 23.54   & -28.24	 & 27.01 & SDSS \\
J090254.16+413506.5 &  09:02:54.17 & +41:35:06.5 & 20.12 & 0.023 & 3.69 & 1.12     & -26.85	 & 25.68 & WHT \\
J090953.85+474943.2 &  09:09:53.85 & +47:49:43.2 & 19.90 & 0.020 & 3.64 & 373.29 & -27.05	 & 28.19 & NOT \\
J091436.23+503848.5 &  09:14:36.23 & +50:38:48.5 & 20.19 & 0.028 & 3.62 & 47.98   & -26.75	 & 27.30 & NOT \\
J091824.38+063653.3 &  09:18:24.38 & +06:36:53.4 & 19.76 & 0.022 & 4.19 & 25.87   & -27.66	 & 27.15 & SDSS \\
J092640.28$-$023041.4 &  09:26:40.29 & $-$02:30:41.5 & 19.82 & 0.021 & 3.76 & 1.90 & -27.20	 & 25.93  & NOT \\
J092832.87+184824.3 &  09:28:32.88 & +18:48:24.4 & 17.54 & 0.016 & 3.77 & 8.67     & -29.49	 & 26.59  & SDSS \\
J093714.48+082858.5 &  09:37:14.49 & +08:28:58.5 & 18.58 & 0.015 & 3.70 & 3.47     & -28.40	 & 26.18  & SDSS \\
J094003.03+511602.7 &  09:40:03.03 & +51:16:02.7 & 19.00 & 0.014 & 3.60 & 12.90     & -27.94	 & 26.72  & SDSS \\
J100012.26+102151.8  & 10:00:12.26 & +10:21:51.9 & 19.54 & 0.023 & 3.64 & 21.87   & -27.41	 & 26.96  & SDSS \\
J101747.75+342737.8 &  10:17:47.76 & +34:27:37.9 & 20.00 & 0.031 & 3.69 & 2.83     & -26.98	 & 26.08  & SDSS \\
J103055.95+432037.7 &  10:30:55.95 & +43:20:37.7 & 19.84 & 0.025 & 3.70 & 31.29   & -27.14	 & 27.13  & SDSS \\
J103420.43+414937.5 &  10:34:20.43 & +41:49:37.5 & 20.12 & 0.029 & 4.00 & 2.17     & -27.12   &  26.03 & NOT \\
J103446.54+110214.4 &  10:34:46.54 & +11:02:14.5 & 18.81 & 0.025 & 4.27 & 1.15     & -28.68	 & 25.81 & SDSS \\
J103717.72+182303.0 &  10:37:17.72 & +18:23:03.1 & 19.81 & 0.023 & 4.05 & 13.59   & -27.48	 & 26.84 & SDSS \\ 
J105121.36+612038.0 &  10:51:21.37 & +61:20:38.1 & 18.92 & 0.020 & 3.69 & 6.90       & -28.06	 & 26.47 & SDSS \\ 
J105756.25+455553.0 &  10:57:56.26 & +45:55:53.1 & 17.45 & 0.022 & 4.14 & 1.10       & -29.93	 & 25.77 & SDSS \\ 
J110147.88+001039.4 &  11:01:47.89 & +00:10:39.4 & 20.18 & 0.026 & 3.69 & 192.10   & -26.79    & 27.92 & SDSS \\ 
J1105:43.86+255343.1 &  11:05:43.87 & +25:53:43.1 & 20.09 & 0.026 & 3.75 & 1.69     & -26.92     & 25.87 & SDSS \\ 
J1109:46.44+190257.6 &  11:09:46.44 & +19:02:57.6 & 20.05 & 0.024 & 3.67 & 7.22     & -26.91    & 26.49 & SDSS \\ 
J1110:55.21+430510.0 &  11:10:55.22 & +43:05:10.1 & 18.58 & 0.024 & 3.82 & 1.21     & -28.50	 & 25.74 & SDSS \\ 
J1117:01.89+131115.4 &  11:17:01.90 & +13:11:15.4 & 18.28 & 0.018 & 3.62 & 28.00     & -28.66	 & 27.07 & SDSS \\ 
J1117:36.33+445655.6 &  11:17:36.33 & +44:56:55.7 & 20.05 & 0.026 & 3.85 & 24.33   & -27.05	 & 27.05 & SDSS \\ 
J1123:39.59+291710.7 &  11:23:39.60 & +29:17:10.8 & 19.46 & 0.017 & 3.77 & 3.68     & -27.57	 & 26.22 & SDSS \\ 
J1125:30.49+575722.7 &  11:25:30.50 & +57:57:22.7 & 19.43 & 0.036 & 3.68 & 2.52     & -27.54	 & 26.03 & SDSS \\ 
J1127:49.45+051140.5 &  11:27:49.45 & +05:11:40.6 & 19.14 & 0.012 & 3.71 & 2.34     & -27.84	 & 26.01 & SDSS \\ 
J1129:38.73+131232.2 &  11:29:38.73 & +13:12:32.3 & 18.78 & 0.026 & 3.61 & 1.77     & -28.16	 & 25.86 & SDSS \\ 
J1133:30.91+380638.1 &  11:33:30.91 & +38:06:38.2 & 19.71 & 0.025 & 3.63 & 1.39     & -27.23	 & 25.76 & SDSS \\ 
\hline\end{tabular}

\medskip
The columns give the following: (1) SDSS object-ID (2) SDSS J2000 coordinates; (3) SDSS dereddened PSF $r$ magnitude; (4) error in PSF $r$ magnitude as given in SDSS;  (5) QSO redshift determined in this work or from SDSS; (6) FIRST peak radio flux density; (7) absolute $r$ magnitude ;  (8) radio luminosity at 1.4 GHz; (9) indicates the source of the data from which the redshift was first obtained (the two WHT sources are from \citealt{Benn2002}).

\end{table*}

\setcounter{table}{3}
\begin{table*}
\caption{(continued)}
\begin{tabular}{r r r r r c r r c r }
\hline
\multicolumn{1}{c}{NAME} &
\multicolumn{1}{c }{RA} &
\multicolumn{1}{c}{DEC} &
\multicolumn{1}{c }{$r_{AB}$} &
\multicolumn{1}{c}{$\sigma_r$} &
\multicolumn{1}{c}{Redshift} &
\multicolumn{1}{c}{$S_{1.4\rm{GHz}}$} &
\multicolumn{1}{c}{$M_r$} &
\multicolumn{1}{c}{log$_{10}P_{1.4\rm{GHz}}$}  &
\multicolumn{1}{c}{ID} \\
\multicolumn{1}{c}{} &
\multicolumn{2}{c}{(J2000)} &
\multicolumn{1}{c}{} &
\multicolumn{1}{c}{} &
\multicolumn{1}{c}{} &
\multicolumn{1}{c}{(mJy)} &
\multicolumn{1}{c}{} &
\multicolumn{1}{c}{(W/Hz)}  &
\multicolumn{1}{c}{}  \\
\multicolumn{1}{c}{(1)} &
\multicolumn{2}{c}{(2)} &
\multicolumn{1}{c}{(3)} &
\multicolumn{1}{c}{(4)} &
\multicolumn{1}{c}{(5)} &
\multicolumn{1}{c}{(6)} &
\multicolumn{1}{c}{(7)} &
\multicolumn{1}{c}{(8)}  &
\multicolumn{1}{c}{(9)} \\
\hline
J113729.42+375224.2 &  11:37:29.43 & +37:52:24.2 & 20.18 & 0.032 & 4.17 & 2.70       & -27.22	 & 26.16 & SDSS \\ 
J115045.61+424001.1 &  11:50:45.61 & +42:40:01.1 & 19.88 & 0.019 & 3.87 & 1.82     & -27.23	 & 25.93  & SDSS \\ 
J115449.36+180204.3 &  11:54:49.36 & +18:02:04.4 & 19.61 & 0.024 & 3.69 & 37.99   & -27.36	 & 27.21  & SDSS \\ 
J120447.15+330938.7 &  12:04:47.15 & +33:09:38.8 & 19.23 & 0.025 & 3.62 & 1.10       & -27.71	 & 25.66 & SDSS \\ 
J121329.42$-$032725.7 &  12:13:29.43 & -03:27:25.7 & 19.64 & 0.027 & 3.67 & 23.37 & -27.33	 & 27.00 & NOT \\ 
J122027.95+261903.5 &  12:20:27.96 & +26:19:03.6 & 18.13 & 0.017 & 3.70 & 34.34   & -28.85	 & 27.17  & SDSS \\ 
J123142.17+381658.9 &  12:31:42.17 & +38:16:58.9 & 20.18 & 0.030 & 4.14 & 20.44   & -27.19	 &   27.03    & SDSS \\ 
J124054.91+543652.2 &  12:40:54.91 & +54:36:52.2 & 19.74 & 0.023 & 3.94 & 14.89   & -27.44	 & 26.86 & SDSS \\ 
J124209.81+372005.6 &  12:42:09.81 & +37:20:05.6 & 19.34 & 0.018 & 3.84 & 644.79 & -27.75	 & 28.47 & SDSS \\ 
J124443.06+060934.6 &  12:44:43.07 & +06:09:34.6 & 19.78 & 0.024 & 3.76 & 1.29     & -27.24	 & 25.76 & NOT  \\ 
J124658.82+120854.7 &  12:46:58.83 & +12:08:54.7 & 20.01 & 0.024 & 3.80 & 1.07     & -27.05	 & 25.69 & SDSS \\ 
J124943.67+152707.0 &  12:49:43.67 & +15:27:07.1 & 19.34 & 0.019 & 3.99 & 1.75     & -27.90	 & 25.94 & SDSS \\ 
J130348.94+002010.5 &  13:03:48.94 & +00:20:10.5 & 18.89 & 0.019 & 3.65 & 1.08     & -28.06	 & 25.66 & SDSS \\ 
J130738.83+150752.0 &  13:07:38.83 & +15:07:52.1 & 19.72 & 0.027 & 4.11 & 3.44     & -27.63	 & 26.26  & SDSS \\ 
J131242.86+084105.0 &  13:12:42.86 & +08:41:05.0 & 18.52 & 0.014 & 3.74 & 4.41     & -28.49	 & 26.29  & SDSS \\ 
J131536.57+485629.0 &  13:15:36.58 & +48:56:29.1 & 19.76 & 0.025 & 3.62 & 9.94     & -27.18	 & 26.61  & SDSS \\ 
J132512.49+112329.7 &  13:25:12.49 & +11:23:29.8 & 19.32 & 0.022 & 4.41 & 69.39   & -28.28	 & 27.61 & SDSS \\ 
J134854.37+171149.6 &  13:48:54.37 & +17:11:49.6 & 19.13 & 0.021 & 3.62 & 2.10       & -27.81	 & 25.94  & SDSS \\ 
J135406.89$-$020603.2 &  13:54:06.90 & $-$02:06:03.2 & 19.17 & 0.018 & 3.72 & 709.05  & -27.82  & 28.49  & SDSS \\
J135554.56+450421.0  &  13:55:54.56 & +45:04:21.1 & 19.36 & 0.021 & 4.09 & 1.48     & -27.98	 & 25.89  & SDSS \\ 
J140635.66+622543.3  &  14:06:35.67 & +62:25:43.4 & 19.73 & 0.020 & 3.89 & 11.03   & -27.41	 & 26.72  & WHT \\ 
J140850.91+020522.7  &  14:08:50.91 & +02:05:22.7 & 19.08 & 0.017 & 4.01 & 1.47     & -28.17    & 25.87  & SDSS \\ 
J142209.70+465932.5  &  14:22:09.70 & +46:59:32.5 & 19.72 & 0.022 & 3.81 & 10.56   & -27.35	 & 26.68  & SDSS \\ 
J142326.48+391226.2  &  14:23:26.48 & +39:12:26.3 & 20.15 & 0.024 & 3.92 & 6.63     & -27.01	 & 26.50  & SDSS \\ 
J143413.05+162852.7  &  14:34:13.06 & +16:28:52.7 & 19.86 & 0.022 & 4.19 & 4.90       & -27.56	 & 26.42 & SDSS \\ 
J143533.78+543559.2 &  14:35:33.78 & +54:35:59.2 & 20.04 & 0.025 & 3.81 & 93.26   & -27.02	 & 27.63  & SDSS \\ 
J144542.76+490248.9 &  14:45:42.76 & +49:02:48.9 & 17.32 & 0.009 & 3.87 & 2.51     & -29.80	 & 26.07  & SDSS \\ 
J144643.36+602714.3 &  14:46:43.37 & +60:27:14.4 & 19.79 & 0.033 & 3.78 & 1.87     & -27.25	 & 25.92  & SDSS \\ 
J145329.01+481724.9 &  14:53:29.01 & +48:17:24.9 & 20.12 & 0.030 & 3.77 & 4.42     & -26.91	 & 26.30  & WHT  \\ 
J150328.88+041949.0 &  15:03:28.89 & +04:19:49.0 & 17.96 & 0.017 & 3.66 & 122.70   & -29.00	 & 27.72  & SDSS \\ 
J150643.80+533134.4 &  15:06:43.81 & +53:31:34.5 & 18.94 & 0.022 & 3.79 & 13.97   & -28.11	 & 26.80  & SDSS \\ 
J151146.99+252424.3 &  15:11:46.99 & +25:24:24.3 & 19.95 & 0.024 & 3.72 & 1.24     & -27.04	 & 25.73  & SDSS \\ 
J152028.14+183556.1 &  15:20:28.14 & +18:35:56.2 & 19.82 & 0.021 & 4.12 & 6.26     & -27.54	 & 26.52  & SDSS \\ 
J153336.13+054356.5 &  15:33:36.14 & +05:43:56.5 & 19.84 & 0.020 & 3.94 & 27.55   & -27.34	 & 27.13  & SDSS \\ 
J161716.49+250208.1 &  16:17:16.49 & +25:02:08.2 & 19.98 & 0.023 & 3.94 & 1.01     & -27.20	 & 25.69  & SDSS \\ 
J161933.65+302115.0 &  16:19:33.65 & +30:21:15.1 & 19.53 & 0.025 & 3.81 & 3.88     & -27.53	 & 26.25  & SDSS \\ 
J163708.29+091424.5 &  16:37:08.30 & +09:14:24.6 & 19.57 & 0.018 & 3.75 & 9.51     & -27.45	 & 26.62  & WHT \\ 
J163950.52+434003.6 &  16:39:50.52 & +43:40:03.7 & 17.95 & 0.017 & 3.99 & 23.78   & -29.28	 & 27.07  & SDSS \\ 
J164326.24+410343.4 &  16:43:26.24 & +41:03:43.4 & 20.10 & 0.025 & 3.87 & 63.40     & -27.02	 & 27.47  & SDSS \\ 
J222814.40$-$085525.7 &  22:28:14.40 & $-$08:55:25.7 & 20.18 & 0.034 & 3.64 & 1.99  & -26.77	 & 25.92  & WHT \\ 
J223535.59+003602.0 &  22:35:35.59 & +00:36:02.1 & 20.15 & 0.027 & 3.87 & 5.06     & -26.97	 & 26.38  & NOT \\ 
J235022.40$-$095144.3 &  23:50:22.40 & $-$09:51:44.4 & 19.67 & 0.021 & 3.70 & 6.37      & -27.31 	 & 26.44  & WHT \\ 
\hline\end{tabular}

\medskip
The columns give the following: (1) SDSS object-ID (2) SDSS J2000 coordinates; (3) SDSS dereddened PSF $r$ magnitude; (4) error in PSF $r$ magnitude as given in SDSS;  (5) QSO redshift determined in this work or from SDSS; (6) FIRST peak radio flux density; (7) absolute $r$ magnitude ;  (8) radio luminosity at 1.4 GHz; (9) indicates where the source redshift was first obtained (the two NED sources are from \citealt{Benn2002}).

\end{table*}

\begin{table}
\centering 
\begin{tabular}{r l}
\hline
\multicolumn{1}{c}{$z_{em}$} &
\multicolumn{1}{c}{$K$ correction} \\
\hline
0.01 \dots\dots & 0.0075\\
0.02 \dots\dots& 0.0151\\
0.03 \dots\dots & 0.0241\\
0.04 \dots\dots & 0.0412\\
0.05 \dots\dots& 0.0720\\
0.06 \dots\dots& 0.1236\\
0.07 \dots\dots& 0.1619\\
0.08 \dots\dots& 0.1851\\
0.09 \dots\dots& 0.1959\\
0.10 \dots\dots&  0.1999\\
0.11 \dots\dots&  0.1985\\
0.12 \dots\dots&  0.1939\\
\hline\end{tabular}
\caption{$K$-correction in the SDSS $r$ band (Fig. 7). A portion of the table is shown here.  The full table is available on-line}
\label{Kcorr2}
\end{table}

\section{Completeness}
Below we estimate the completeness (number of QSOs selected divided by the actual number of QSOs in this range of redshift)  of the sample of 87 radio-loud QSOs with $3.6 \le z \le 4.4$ .  This sample will be used (in Section 7) to calculate the luminosity function. There are several sources of incompleteness: exclusion of optical images of poor quality (Section 6.1); radio-survey incompleteness and missed radio-optical identifications (Section 6.2); and incompleteness of selection by the neural-network algorithm (Section 6.3).

\subsection{Optical image quality}

Due to the sensitivity of our NN to data of poor photometric quality, 
we discarded at the pre-selection stage (Section 2.2) 
objects having 'fatal' error flags or magnitude errors larger than 0.2 in all five bands (Richards et al. 2002) or flagged CHILD.

Incompleteness due to the exclusion of fatal and non-fatal photometric errors ( \citealt{Richards:2002tg} define as 'non-fatal' errors: some empirical combination of SDSS flags generally associated with poor de-blends of complex objects) during the SDSS selection of QSOs-candidates, was discussed by \cite{Richards:2006ye}  who applied a global 5\% correction. In previous evaluations of this selection effect,  (\citealt{Croom:2004ij})  suggest 6\% incompleteness for objects with $17.5 \le i \le 18.5$, and \cite{Vanden-Berk:2005kx} estimated an incompleteness of 3.8\% for point-like objects with $i<19.1$ . 

To quantify the incompleteness of our selection due to the exclusion
of CHILD objects, we evaluated the fraction of such objects amongst
$15 \le r \le 20.2$ QSOs in the 5th SDSS Quasar Catalogue, which did not
exclude CHILD objects. The fraction that we derive in this way is 27\%.
The net completeness due to exclusion of these two types of object is therefore 
69\% ($0.95 \times 0.73$).

These fractions are in approximate agreement with the statistics of
Table 1, which indicate 4.8\% incompleteness due to 'fatal' errors
(669 rejected out of a total of 13956 sources), and 31\%
incompleteness due to the exclusion of sources flagged as CHILD (4148
out of 13287), i.e. a net completeness of 66\% ($0.95 \times 0.69$).

For the analysis here, we adopt an intermediate estimated completeness of 68\%.

 We do not apply any correction for objects misclassified in SDSS as
 having galaxy morphology but being star-like objects. 
Tests of random samples of sources with available spectra indicate
that this source of incompleteness is $< 0.03\%$.

\subsection{Radio incompleteness}
Two sources of incompleteness arise from radio selection using the FIRST survey. 

\subsubsection{Incompleteness of FIRST survey}

The completeness of the FIRST survey as a function of flux density has
been estimated for SDSS quasars and is given in fig. 1 of
\cite{Jiang:2007mi} (and is discussed elsewhere, e.g. 
by \citealt{Prandoni:2001kl}).  
At integrated FIRST flux density $>3$ mJy FIRST
is $>96\%$ complete and the completeness declines with decreasing flux density.
From the results of these studies we assume the following
completeness function, $q(S)$, where $S$ is the FIRST integrated flux
density in mJy:
 
\begin{equation}
q(S) =
\begin{cases}
0.50, & S\leq 1.25\\
0.75, & 1.25 <  S \le 2 \\
0.85, & 2 < S \le 3\\
0.95, & 3 <  S \le 5\\
1, & S > 5
\end{cases}
\label{qs}
\end{equation}

\begin{figure}
\centering
\includegraphics[width=0.5 \textwidth]{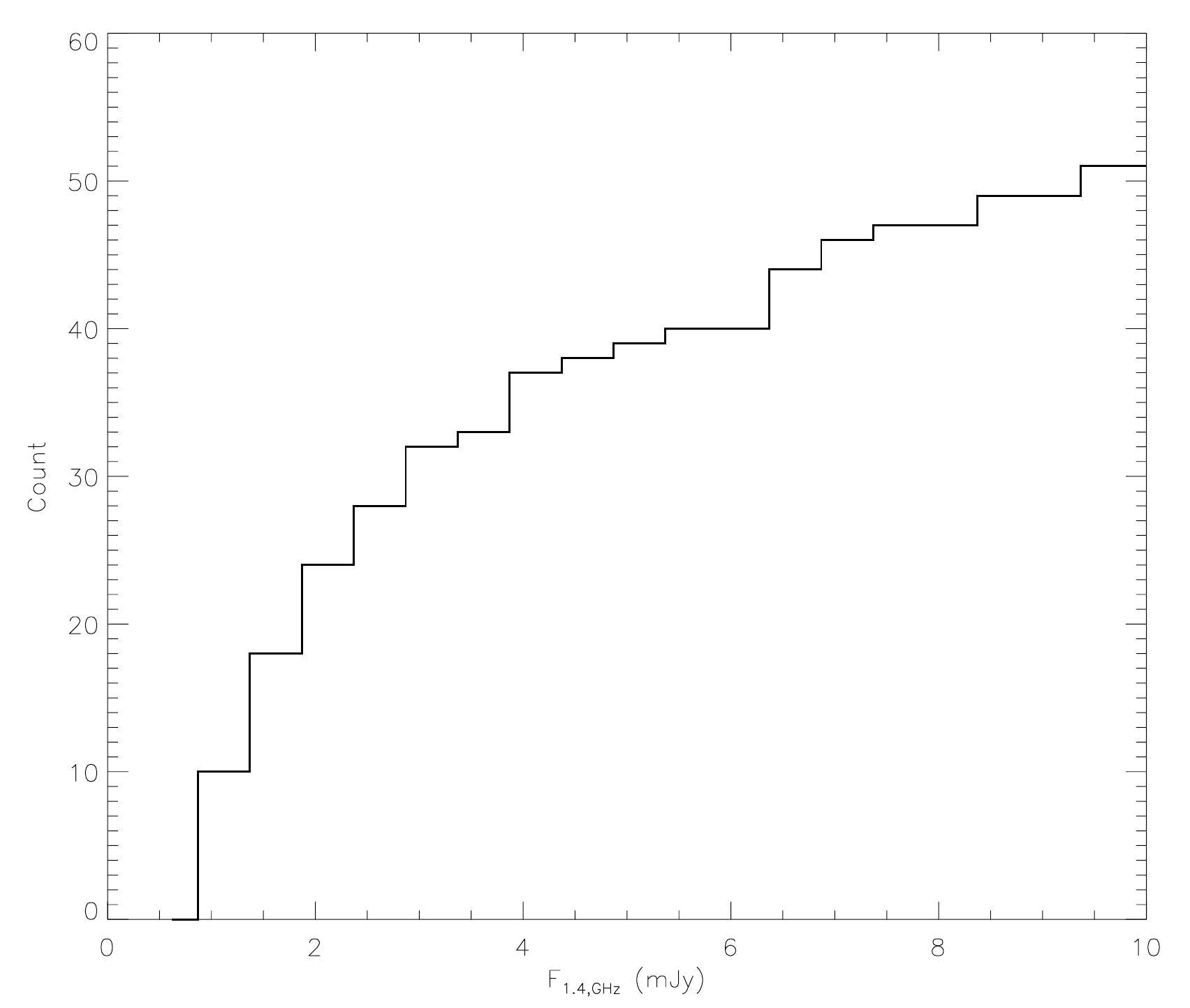}
\caption{Cumulative distribution of FIRST flux densities 
for those QSOs in our sample of 87 with $3.6 \le z \le 4.4$,
and with radio flux density $<$ 10 mJy.}
\label{IntegratedFlux}
\end{figure}

 In Fig. \ref{IntegratedFlux} we show for our sample the cumulative distribution of integrated radio flux densities, for objects fainter than 10 mJy.

Applying $q(S)$ to our final sample we obtain a completeness of 84 per cent (87 QSOs detected and  $\sim 16$ missed), for the flux density limit of 1 mJy. 

\subsubsection{Match radius}
To obtain our radio-optical sample we sought simple one-to-one matches
within radius 1''.5.  We used this criteria for consistency with C08,
who adopted this on the basis that more than 99\% of FIRST-APM QSOs
with $3.8\le z \le 4.5$, $E\le 18.8$ and $S_{1.4GHz} >1$ mJy fall
within this radius \cite{Vigotti:2003vv}.

However, a simple one-to-one match between FIRST and SDSS will miss double-lobe QSOs without detected radio cores. De Vries, Becker \& White (2006) found that for a sample of $5,515$ FIRST-SDSS QSOs with radio morphological information within 450'', the fraction of FIRST-SDSS double-lobe QSOs with undetected cores 
is $3.7$\%. Since the starting samples of SDSS QSOs in \citep{de-Vries:2006ly}  and in this work obey similar SDSS selection criteria, we used this value to correct for this source of incompleteness.

\subsection{Incompleteness of selection by the neural-network}

The completeness of our neural network classifier was estimated as 97
per cent, from the testing on known high-z QSOs presented in Section
3.3. The classifier selected 15 high-z QSO candidates and rejected
2901.  Of the later, 451 have now spectroscopy from DR9 and none of
them was identified as a high-z QSO, confirming this high level of
completeness (see Section 4.2).

\subsection{Net completeness}
The net completeness of our sample of 87 RL-QSOs at $3.6\le z \le 4.4$
(Section 5), is the product of each of the completeness terms
discussed above:
acceptance of only those candidates with high-quality photometry (Section 6.1, 68\% completeness);
completeness of the FIRST radio survey (Section 6.2.1, 84\% for our sample);
acceptance of only optical-radio matches within a given radius and 
exclusion of extended sources (Section 6.2.2, 99\% and 96.3\%); 
and the completeness of the NN selection algorithm (Section 6.3, 97\%).

Multiplying these four terms together, the net completeness for our sample is 53\%.

\section{Binned Luminosity function}
\subsection{Method}

Using the final sample of 87 QSOs listed in Section 5 and correcting for incompleteness as discussed in Section 6, we compute the binned Quasar Luminosity Function (QLF)  in the redshift range  $3.6 \le z \le 4.4$.  

The binned QLF is usually calculated using the classical $1/V_{max}$
method (\citealt{Schmidt:1968kl}; \citealt{Maccacaro1991} ;
\citealt{Ellis1996}), or its generalized version (usually known as $
\sum V_a^{-1}$) applied to samples comprising subsamples with
different flux limits (\citealt{Avni:1980dq}).  

The $V_{max}$ method is an unbiased \citealt{Felten1976}) non-parametric
estimator of the space density.  It is commonly used to fit models of
the LF, since it has the advantage that it does not assume any
underlying model. Even when the model LF is fitted to the unbinned
data (for example in the maximum likelihood technique of 
\citealt{Marshall1983}) it is often used before performing the fit to
observe the overall behavior of the LF.

However \cite{Page:2000vn} demonstrated that the $\sum 1/V_{max}$
estimator introduces significant errors for objects close to the flux
limits of the survey. An alternative method proposed by
\cite{Page:2000vn} is superior and partially corrects for this source
of error, although implicitly assumes a uniform distribution of the
sources within each bin (\citealt{Croom2009} ; \citealt{Miyaji2001}).
The variation of the LF within a bin can be particularly critical at
the steep bright end of the QSO LF.  Instead, we used a modified
version of the \cite{Page:2000vn} method that does not make use of the
uniform-distribution assumption and is still model-independent.

To illustrate the method used in this paper, we start with a brief overview of the $1/V_{max}$ and the \cite{Page:2000vn} methods. The luminosity function is defined as the number of objects per unit of comoving volume, per unit of luminosity. A naive approach to the calculation of space density in an interval $[L_1,L_2] \times  [V_1(z_1), V_2(z_2)]$ of luminosity and redshift, centered upon values $L^*$ and $z^*$,  would be to simply count the number of objects $N$ within the interval considered:

\begin{equation}
\Phi (L^*,z^*) = \frac{N}{\Delta V \Delta L}  
\end{equation}

\noindent
The $1/V_{max}$ method, first proposed by  \cite{Schmidt:1968kl}  takes into account the fact that in flux-limited samples there is a higher probability to observe a  bright source than a faint one. Thus, the count of sources $N$ is replaced with a sum of probabilities: 

\begin{equation}
\sum_1^N \frac{V_0}{V_{max,i}} 
\end{equation}

\noindent
 where $V_0$ is the volume  over which  we are computing the luminosity function,and $V_{max,i}$ is the maximum volume at which the source could be observed and still be included in  our sample. In this way, the computation of the LF becomes
 
\begin{equation}
\Phi (L,z) = \frac{1}{\Delta L}  \sum_1^N \frac {1} {\Delta V_{max,i}}
\end{equation}

\noindent
\cite{Page:2000vn} noted that the limit in apparent magnitude of the survey bounds the region of integration. In particular, for a given bin in redshift, $L_1$ and $L_2$ should be replaced by the actual luminosity limits ($L_{min}, L_{max}$) as determined by the intersections with the limiting-magnitude curves of the survey.
Therefore,  $\Phi (L,z)$ is calculated as 

\begin{equation}
\Phi(L,z)=\frac{N}{\int_{L_{min}}^{L_{max}} \int_{z_{min}}^{z_{max(L)}} (dV/dz) dz \, dL}
\end{equation}

\noindent
where $z_{min}$ is the bottom of the redshift interval and $z_{max}(L)$ is the highest possible redshift for an object of luminosity L within the considered bin $\Delta z$. This approach takes into account the real area of integration but implicitly assumes uniform distribution of sources over the bin. 

In order to minimise this bias, we calculate the maximum actual integration area determined as in \cite{Page:2000vn} but for each source in the bin, so as not to lose the $V_{max}$ information for individual sources. Then we sum over all the sources in the bin. In this way we do not count the number of sources over an area larger 
than that of the actual survey and at same time, within a single bin,  we weight  sources by luminosity.  Finally,  $\Phi (L,z)$ is calculated as

\begin{equation}
\Phi (L,z) =  \sum_{i=1}^N \frac{1}{\int_{L_{min}}^{L_{max,i}} \int_{z_{min}}^{z_{max(L),i}} (dV/dz) dz \, dL} 
\end{equation}

\begin{figure*}
\centering
\includegraphics[width=1 \textwidth]{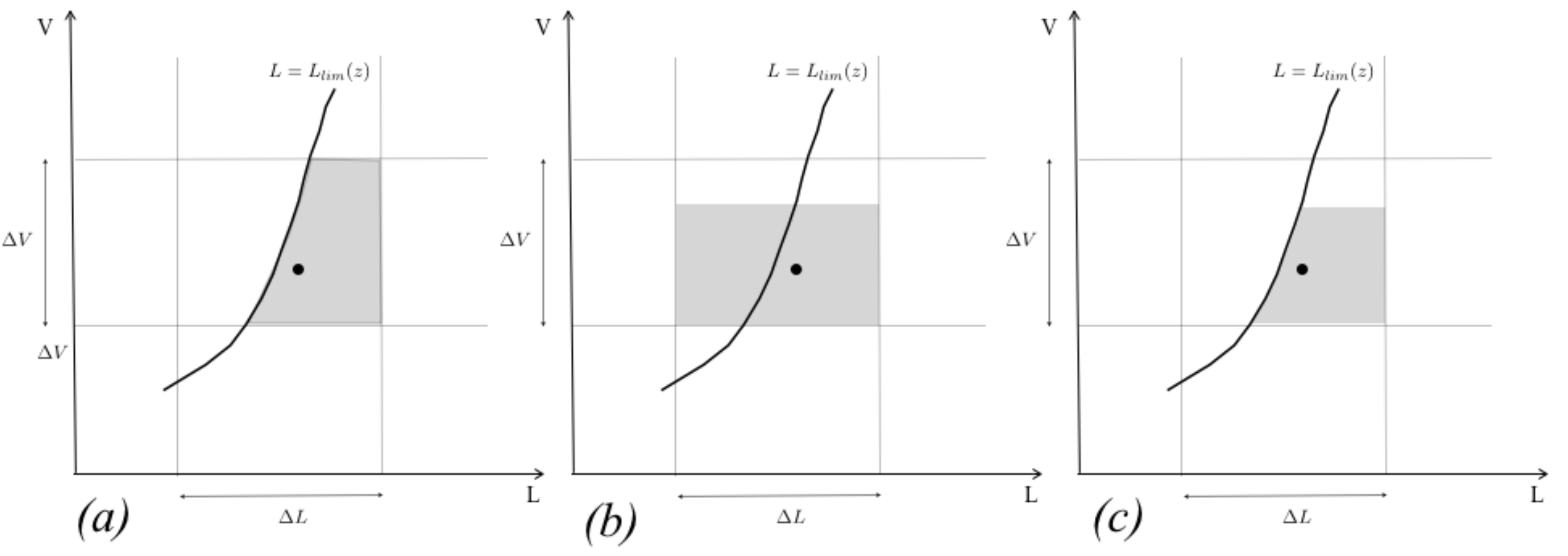}
\caption{Volume-luminosity space used to calculate the contribution to the LF from a single object (the black dot) in a given bin intersected by the line $L = L_{lim}(z)$, i.e. the minimum detectable luminosity of an object at redshift $z$.  The available space (grey shaded area) is shown for a binned LF calculated using: (a) the Page \& Carrera implementation , (b) the classical $1/V_{max}$ method, (c) our methodology.}
\label{comparisonMethodsLF}
\end{figure*}

The difference between this methodology and that of \cite{Page:2000vn}
is negligible in the case of a large sample of QSOs uniformly
distributed in (M,z) space, but becomes critical for small samples not
distributed uniformly in each bin, as is the case here.  In
Fig. \ref{comparisonMethodsLF} we show the volume-luminosity space
available to an object in a bin intersected by a
limiting-magnitude curve, (a) in the \cite{Page:2000vn} approach, (b) in
the classical $1/V_{max}$ case, and (c) for the methodology used here.

The statistical uncertainty $\delta \Phi$ is calculated for each bin i as 
\begin{equation}
\delta \Phi_i = \frac{\Phi_i}{\delta N_0}
\end{equation}

Where  $N_0$ is the actual number of objects in the bin and $\delta N_0 = \sqrt{N_0}$ (Poisson statistics).   
The formula is easily derived, when the space density $\Phi_i$  is assumed to be approximately:

\begin{equation}
\Phi_i \approx \frac{N}{V_e  L_e} = \frac{N_0 \cdot f}{V_e  L_e} 
\end{equation}

where $(V_e  L_e)$ is an equivalent space-luminosity area, and $N$ is the corrected number of quasars 
in bin i.  The error in the completeness factor $f$ is assumed to be $\approx 0$.

\subsection{The QSO Luminosity Function}

\begin{figure*}
\centering
\includegraphics[width=0.6  \textwidth,angle=270]{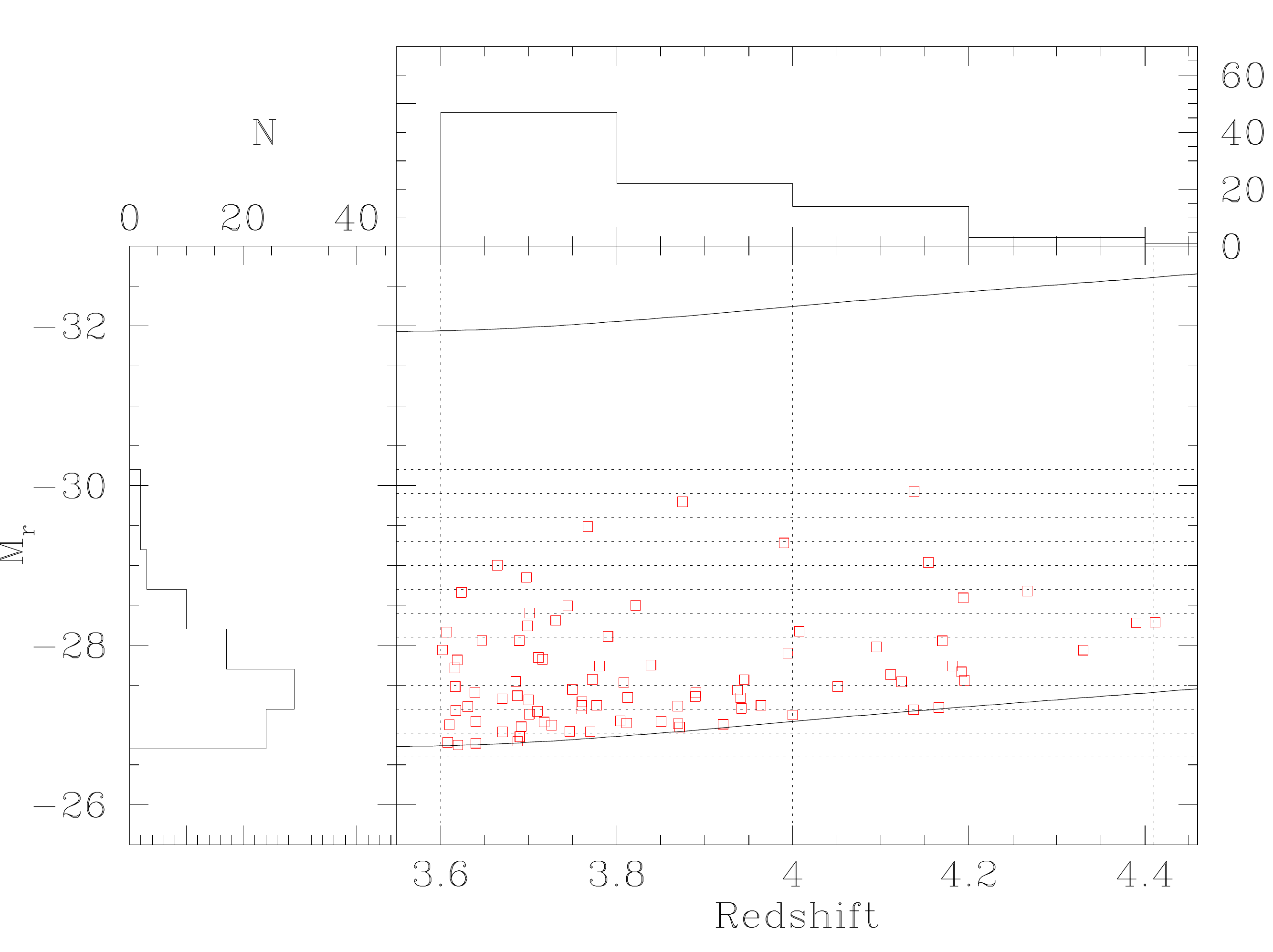}
\caption{Luminosity-redshift diagram for the complete sample of 87 radio loud QSOs. The dotted lines show the limits of the bins ($\Delta L\Delta z$) used to compute the luminosity function. Solid lines represent the upper ($r=15$) and lower ($r=20.2$) limits of the survey. Upper and left panels show marginal histograms of redshift and absolute magnitude, respectively (see Section 7.2)}
\label{Mvsz}
\end{figure*}

The QLF was computed using  two bins in redshift, 3.6 - 4.015 and 4.015 - 4.415, and 11 bins in  
optical absolute magnitude starting with $M_r = -26.6$ and with $\Delta M = 0.3$. In Fig. \ref{Mvsz} we plot  $M_r$ vs. redshift for our sample of QSOs; the dotted grid shows the bins in magnitude and redshift used to compute the QLF. The curves show the upper and lower limiting apparent magnitude r of our selection. The top and side panels show the marginal distributions in redshift and absolute magnitude, respectively.

Since our complete sample results from  two  surveys with different flux limits, the maximum redshift at which a source can be observed  may be different in the two surveys.
The most efficient way to combine areas with different flux limits is to assume that each object could be found in any of the survey areas for which it is brighter than the corresponding flux limit. This is `coherent' addition of samples, in the language of \cite{Avni:1980dq} .
Therefore, since our survey is a radio-optical survey, for each source we chose the smaller of $z_{max, optical}$ and  $z_{max, radio}$. 

\begin{figure}
\centering
\includegraphics[width=0.48  \textwidth]{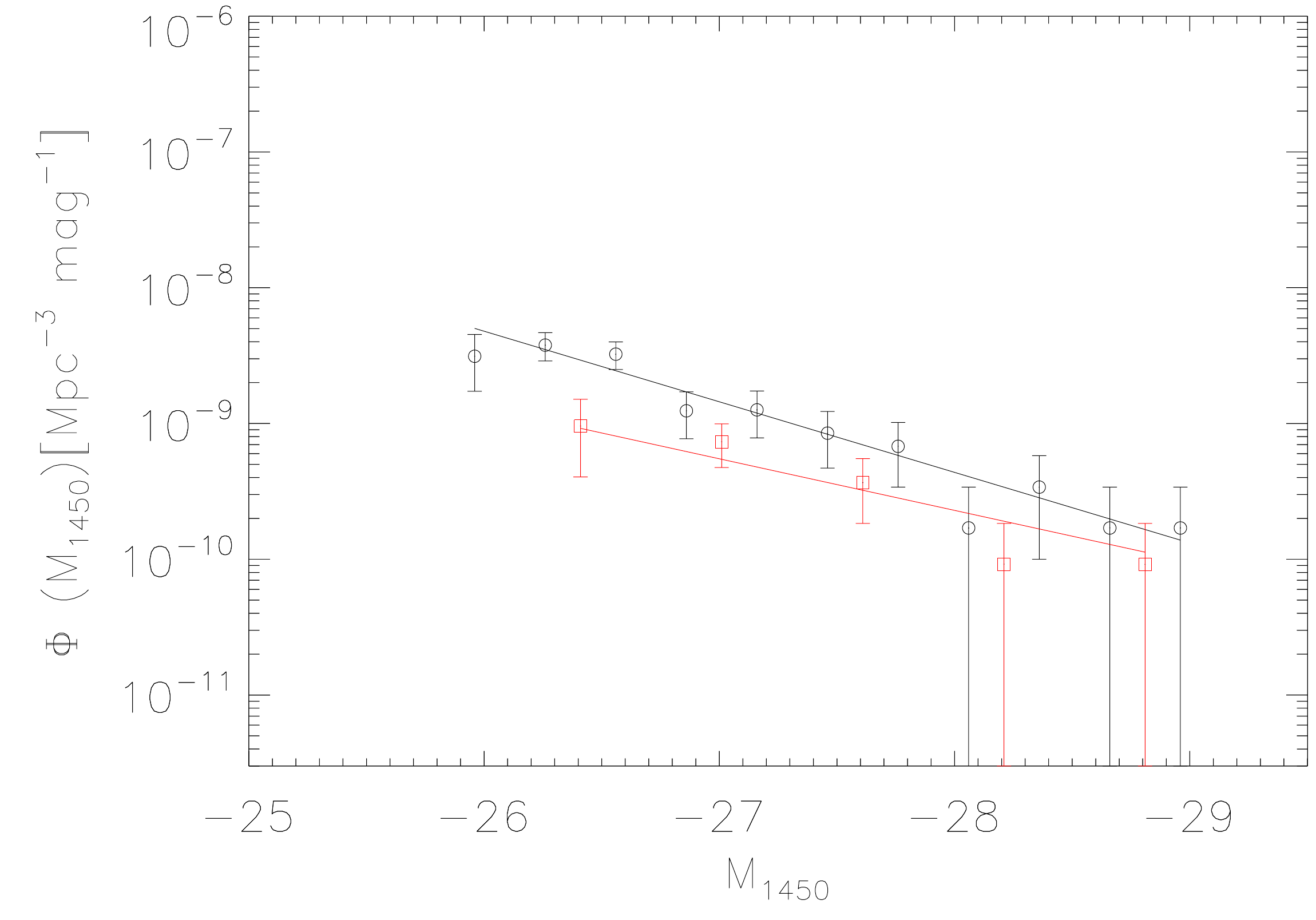}
\caption {The luminosity function derived from our sample of 87  radio-loud QSOs with ${\rm log}_{10} P_{1.4, \rm GHz} {\rm (W/Hz)} > 25.7$ . The luminosity function for $z \sim 3.8$ is shown with black points, while red squares show the luminosity function for  $z\sim 4.2$. Black and red lines are the best fit-slopes.}
\label{figure10e5}
\end{figure}

For each bin we applied the completeness corrections as explained in Section 6 and computed the weighted number of QSOs in the bin. Table \ref{QLF} and Figure \ref{figure10e5} show the resulting QLF for radio-loud QSOs. In Fig. \ref{figure10e5} we plot separately the luminosity function for the two bins of redshift, at $z\sim 3.8$, and $z\sim 4.2$. We also show the best-fit slopes which will be discussed in Sect. 8.3.

\begin{table*}
\caption{Binned luminosity function for FIRST-SDSS quasars at redshift $\sim 4$}
\begin{tabular}{c r r c c c c r r c}
\hline
\hline
\multicolumn{1}{c}{Redshift} &
\multicolumn{1}{c}{$M_r$} &
\multicolumn{1}{c}{$\phi_{RL} (\times 10^{-9}$)} &
\multicolumn{1}{c}{$\sigma_\phi (\times 10^{-9}$)} &
\multicolumn{1}{c}{$RLF$} &
\multicolumn{1}{c}{$\phi (\times 10^{-8}$)} &
\multicolumn{1}{c}{fill} &
\multicolumn{1}{c}{$N_Q$} &
\multicolumn{1}{c}{$N_{Qcorr}$} & 
\multicolumn{1}{c}{$N_r$} \\
\multicolumn{1}{c}{(1)} &
\multicolumn{1}{c}{(2)} &
\multicolumn{1}{c}{(3)} &
\multicolumn{1}{c}{(4)} &
\multicolumn{1}{c}{(5)} &
\multicolumn{1}{c}{(6)} &
\multicolumn{1}{c}{(7)} &
\multicolumn{1}{c}{(8)} &
\multicolumn{1}{c}{(9)} &
\multicolumn{1}{c}{(10)} \\
\hline
$3.8$ & -26.75 &  3.129 & 1.400  & 3.17($\pm 1.60$)\%   & 9.870  & 1  & 5 & 9.43  & 0   \\
$3.8$ & -27.05 & 3.778  & 0.890  & 3.58($\pm 1.82$)\%   &10.559 & 1  & 18 & 33.96 & 0 \\
$3.8$ & -27.35 & 3.242  & 0.744  & 4.04($\pm 2.07$)\%   & 8.025  & 0  & 19 & 35.85  & 1  \\
$3.8$ & -27.65 & 1.240  & 0.469  & 4.56($\pm 2.36$)\%   & 2.717  &  0  & 7 & 13.21  & 1  \\
$3.8$ & -27.95 & 1.260  & 0.476  & 5.15($\pm 2.69$)\%   & 2.447  &  0  & 7 & 13.21  & 1  \\
$3.8$ & -28.25 & 0.849  & 0.379  & 5.80($\pm 3.01$)\%   & 1.463  &  0  & 5 & 9.43  & 0  \\
$3.8$ & -28.55 & 0.679  & 0.339  & 6.53($\pm 3.50$)\%   & 1.040  &  0 & 4 & 7.55  & 0  \\
$3.8$ & -28.85 & 0.170  & 0.170  & 7.35($\pm 4.00$)\%   & 0.231  &  0 & 1 &  1.89 & 0  \\
$3.8$ & -29.15 & 0.340  & 0.240  & 8.26($\pm 4.58$)\%   & 0.411  &  0 & 2 & 3.77  & 0  \\
$3.8$ & -29.45 & 0.170  & 0.170  & 9.26($\pm 5.24$)\%   & 0.183  &  0 & 1 & 1.89  & 0  \\
$3.8$ & -29.75 & 0.170  & 0.170  & 10.39($\pm 6.01$)\% & 0.163  &  0 &  1 & 1.89   & 0  \\
          &            &              &           &               &           &     &     &           &    \\
$4.2$ & -27.05 & 1.920 & 1.920    & 3.04($\pm 1.60$)\%   & 6.316 &  1 & 1 & 1.89  & 0  \\
$4.2$ & -27.35 & 0.657 &  0.464   & 3.44($\pm 1.81$)\%   & 1.910 &  1 & 2 & 3.77  & 0  \\
$4.2$ & -27.65 & 0.917 & 0.410    & 3.89($\pm 2.07$)\%   & 2.358 &  0 & 5 & 9.43  & 0  \\
$4.2$ & -27.95 & 0.550 &  0.318   & 4.39($\pm 2.36$)\%   & 1.254 &  0 & 3 & 5.66  & 0  \\
$4.2$ & -28.25 & 0.367 &  0.260   & 4.95($\pm 2.69$)\%   & 0.741 &  0 & 2 & 3.77  & 0  \\
$4.2$ & -28.55 & 0.367 &  0.260   & 5.58($\pm 3.07$)\%   & 0.658 &  0 & 2 & 3.77  & 0  \\
$4.2$ & -29.15 & 0.183 &  0.183   & 7.07($\pm 4.01$)\%   & 0.260 &  0 & 1 & 1.89   & 0  \\
$4.2$ & -30.05 & 0.183 &   0.183  & 10.02($\pm 6.03$)\% & 0.183 &  0 & 1 & 1.89  & 0  \\
\hline
\end{tabular}
\label{QLF}

\medskip
The columns give the following: (1) median redshift of the bin, (2) 
median absolute magnitude $M_r$ of the bin, (3) space density $\phi_{\rm RL}$ (${\rm Mpc}^{-3}$  ${\rm mag}^{-1} (\times 10^{-9}$)) of the radio-loud QSOs , (4) error on the space density, $\sigma_\phi (\times 10^{-9}$) , (5)  radio-loud fraction calculated using \cite{Jiang:2007mi},  (6) ) space density $\phi$ ($Mpc^{-3}$  $mag^{-1} (\times 10^{-8}$)) of the QSOs (RQ+RL), (7) indication if the bin is intersected by the limiting magnitude curve (1 = yes and 0 = not) , (8) actual number of QSOs in the bin, (9) the corrected number of QSOs in the bin after applying the completeness corrections, (10) Number of QSOs in the bin limited by the radio-flux limit.
 \end{table*}

\section{Discussion}

Below we use the binned LF calculated from the previous section to
derive the space density of RL QSOs at $3.6 \le z \le 4.4$ (Section
9.1). We then derive the space density of the entire population (i.e
radio-loud + radio-quiet) of QSOs at this redshift (Section 8.2) using
two independent estimates of the radio-loud-fraction (Sections 8.2.1,
8.2.2). Finally we derive the slope of the LF of RL QSOs at $z \sim
3.8$ and at $z \sim 4.2$, and compare our results with those of other
authors (Section 8.3).

Our sample includes QSOs with optical luminosities $M_r < -26.6
\Leftrightarrow M_{1450} \lesssim -25.8$ (see equation
\ref{conversione}). 
We adopt the definition of radio-loudness 
used by \cite{Gregg1996}, i.e. log$ \, P_{1.4, \rm GHz} >
25.5$.  All the sources included in our sample (i.e. $S_{1.4\rm GHz} >
1 $\ mJy) meet this criterion and are therefore 
radio-loud. Due to the limit in flux density, the minimum radio luminosity of
sources included in our sample depends on redshift. In particular,
using a radio spectral index of $\alpha_r =-0.3$, at the lowest
redshift of our sample, i.e. $z=3.6$, the FIRST flux-density limit
corresponds to radio luminosity log$\, P_{1.4, \rm GHz} (W/Hz) >
25.61$.  For $z= 4$, it corresponds to log$\, P_{1.4, \rm GHz} (W/Hz) >
25.7$ and for $z = 4.4$ to log$\, P_{1.4, \rm GHz} (W/Hz) > 25.77$.

\subsection{The space density of RL QSOs at $3.6 \le z \le 4.4$}

Starting with the binned luminosity function determined in Section 8.2 we calculate the space density of RL QSOs with optical luminosity $M_{1450} \lesssim  -25.8$ and  radio luminosity log$\, P_{1.4, \rm GHz} (W/Hz) > 25.5$, in two shells of redshift. The first shell has median $z \approx 3.8$  ( $3.6 \le z \le 4.015$) and the second shell  has median $z \approx 4.2$ ($4.015  \le z \le 4.415$). Integrating the binned LF (Table 6), the space densities of QSOs  are therefore:

\begin{align*}
\rho(z \approx 3.8,M_{1450} < -25.8)_{\rm RL} = 4.51 \pm 0.61  \, {\rm Gpc}^{-3}  \\
\rho(z \approx 4.2,M_{1450} < -25.8)_{\rm RL} = 1.54 \pm 0.63  \, {\rm Gpc}^{-3} 
\end{align*}

\begin{figure}
\centering
\includegraphics[width=0.5 \textwidth]{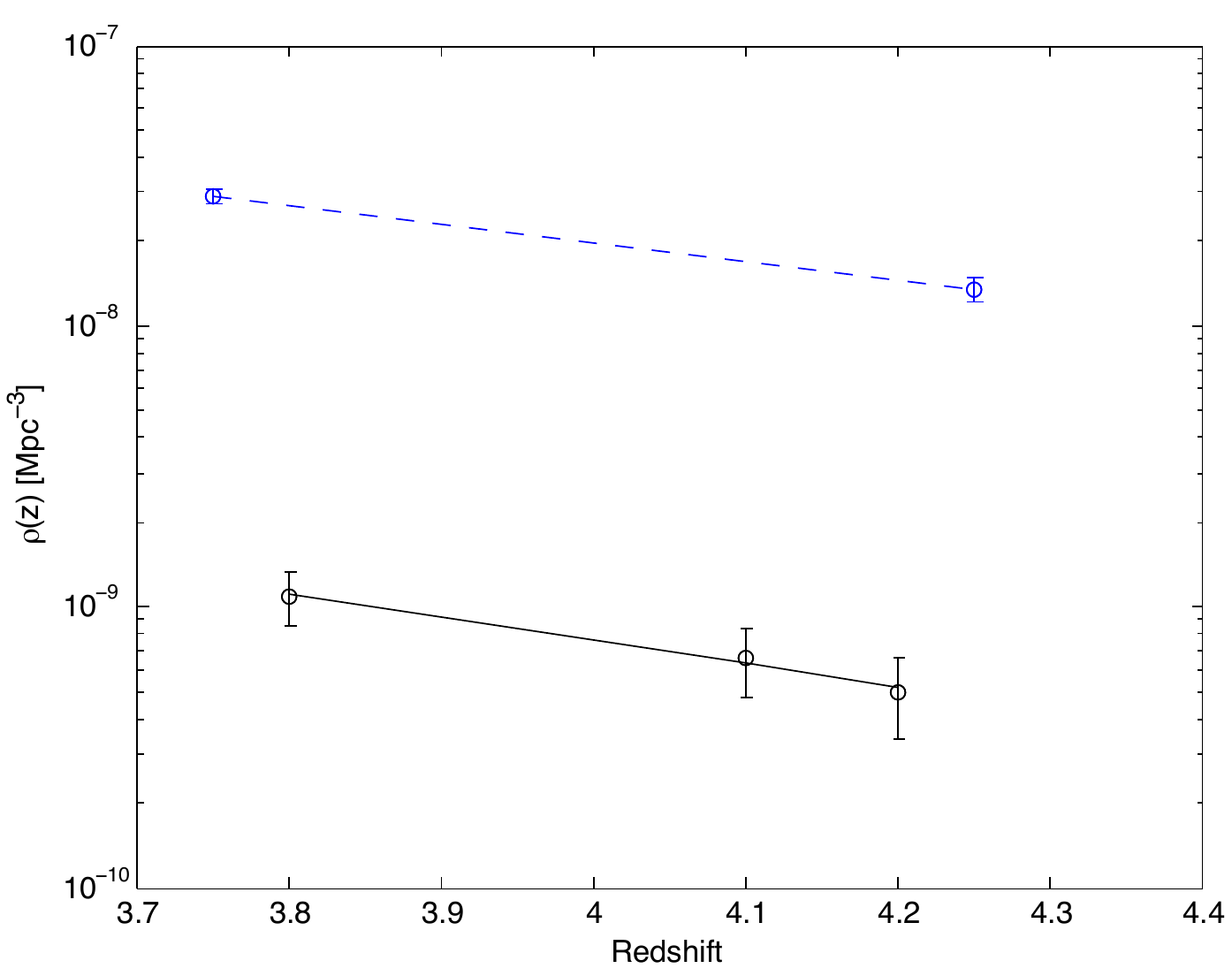}
\caption{The black points show the integrated luminosity function for RL QSOs with $M_{1450} < -27.0$. The points for $z \sim 3.8$ and $z \sim 4.2$ were obtained in this work, showing good agreement with the point obtained for $z \sim 4.1$  in Vigotti et. (2003). For comparison, blue points show the space density for the entire population of QSOs and for  $M_{1450} <-27.6$, as found in Richards et al. (2006). See Section 8.1.}
\label{spaceDensityZ}
\end{figure}

From a sample of radio QSOs obtained by cross-matching the FIRST radio survey and the Automatic Plate Measuring Facility catalogue of POSS I, \cite{Vigotti:2003vv} measured the space density at $ 3.8 \le z \le 4.5$ of optically-luminous ($M_{1450} < -26.9$) radio-loud QSOs and obtained  $\rho(z \approx 4.1 ,M_{1450} < -26.9)_{\rm RL}= 0.99 \pm 0.28  \, {\rm Gpc}^{-3}$. We recalculated the space density and optical luminosities using our adopted cosmology (noted at the end of Section 1), obtaining  $\rho(z \approx 4.1 ,M_{1450} < -27.1)_{\rm RL}= 0.66 \pm 0.18  \, {\rm Gpc}^{-3}$.
By integrating our binned LF in the interval $M_{1450} \lesssim -27.0$,  we obtain:

\begin{align*}
\rho(z \approx 3.8,M_{1450} < -27.0)_{\rm RL} = 1.09 \pm 0.24  \, {\rm Gpc}^{-3}  \\
\rho(z \approx 4.2,M_{1450} < -27.0)_{\rm RL} = 0.50 \pm 0.16  \, {\rm Gpc}^{-3} 
\end{align*}
in good agreement with \cite{Vigotti:2003vv}  (see Fig. \ref{spaceDensityZ}) and consistent with a linear decrease of space density with increasing redshift.  

Using a sample of QSOs obtained by cross-matching FIRST and SDSS-DR6, \cite{McGreer:2009nx} calculated  a binned luminosity function in the redshift range $3.5 \le z \le 4.0$. These authors used the same starting surveys as we did and a similar range of redshift, but they calculated the LF only for QSOs with radio-loudness parameter $R>70$. The $R$ parameter is another common criterion for distinguishing between radio-quiet  and radio-loud AGN.  It is defined (\citealt{Kellermann1989}; \citealt{Stocke1992}) as the rest-frame ratio of the  monochromatic 6-cm (5 GHz) and 2500\AA \, flux densities. Generally, objects are considered to be RL for $R>10$. 

The space density calculated in \cite{McGreer:2009nx} for $M_{1450} < -26.1$ is  $\rho(z = 3.75,M_{1450} < -26.1)_{\rm R>70} = 1.38 \pm 0.59  \, {\rm Gpc}^{-3}$. The cosmology parameters used by \cite{McGreer:2009nx} are the same that we use. 

At redshift $\approx 4$ our definition of radio-loudness is very close
to the common definition $R>10$, but we needed to re-calculate the LF
using a subsample of RL-QSOs with $R>70$ in order to compare our LF
with \cite{McGreer:2009nx}. To calculate the $R$-parameter for our
sample of QSOs, we used $\alpha_\nu = -0.5$ (in agreement with
\citealt{McGreer:2009nx}) to transform the flux from $S_{1.4 \rm GHz}$
to $S_{5 \rm GHz}$. We follow \cite{Oke1983} when converting from
magnitude to luminosity ($2500$ \AA). In this way, we obtain
\begin{equation}
\rho(z = 3.8,M_{1450} < -26.1)_{\rm R>70} = 2.49 \pm 0.36 \, {\rm Gpc}^{-3}
\end{equation}
which is a factor $1.8$ ($2\sigma$) higher than the value $\rho= 1.38
\pm 0.59 \, {\rm Gpc}^{-3}$ found by \cite{McGreer:2009nx}. 
This difference may in part be ascribed to the higher completeness of our NN selection,
and in part to the smaller FIRST-SDSS matching radius used by
\cite{McGreer:2009nx}, which will exclude some quasars.

In Fig. 13 we show the cumulative luminosity functions for the two redshift bins (i.e. $z = 3.8$ and $z = 4.2$). Each point of the cumulative function is the space density $\rho(< M_{\rm 1450})$ as a function of absolute magnitude. The two functions can be compared with previous results by \citet{Vigotti:2003vv} and \citet{Carballo:2006uq}, at redshifts $z \sim 4.1$ and $z\sim 4$, respectively. As expected, due to the evolution of space density with redshift, these last two values lie between our determinations. 

\begin{figure}
\centering
\includegraphics[width=0.5 \textwidth]{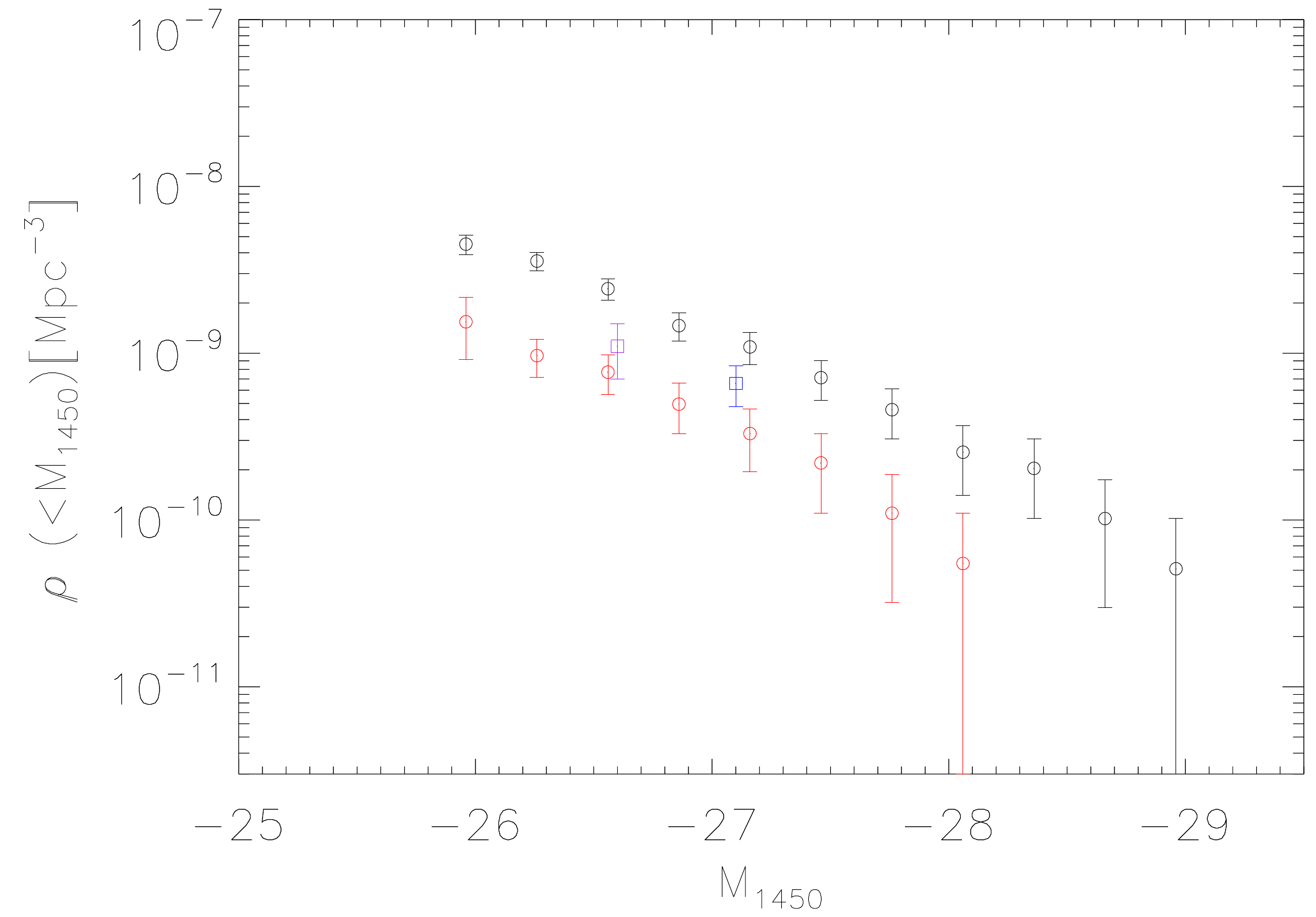}
\caption{Cumulative luminosity functions at $z=3.8$ (black circles) and $z=4.2$ (red circles). Squares represent densities derived by \citet{Vigotti:2003vv} (blue square) and \citet{Carballo:2006uq} (purple square) at redshifts $z\sim 4.1$ and $z\sim 4$, respectively.}
\label{cumulative}
\end{figure}

\subsection{Total space density of QSOs at $3.6 \le z \le 4.4$}

From the space density of the radio-loud QSO population we can roughly
test the predictions of Radio-Loud-Fraction (RLF) over this range of
redshift, by comparing the implied total space densities with measurements 
of space density from the literature.

  It has long been known   that between 5\% and 15\% of all quasars are radio-loud (e.g., \citealt{Kellermann1989}; \citealt{Urry1995}; \citealt{Ivezi2002}). 
However, some authors conclude that the RLF does not change significantly with redshift (e.g., \citealt{Goldschmidt1999}; \citealt{Stern2000}; \citealt{Cirasuolo:2003qf}) or luminosity (e.g., \citealt{Bischof1997}; \citealt{Stern2000} \citealt{Vigotti:2003vv}), while others find that the RLF decreases with increasing redshift (e.g., \citealt{Peacock1986} ;\citealt{Miller1990}; \citealt{Visnovsky1992}; \citealt{Schneider1992}) and decreasing opical luminosity (\citealt{Padovani1993}; \citealt{LaFranca1994}); or that it evolves non-monotonically with redshift and luminosity (e.g. \citealt{Hooper1995}).

We therefore derive below the space density of all QSOs in two different ways: assuming constant RLF (Section 8.2.1); and using a recently-determined redshift- and luminosity- dependent RLF  (\citealt{Jiang:2007mi}) (Section 8.2.2).

\subsubsection{For constant RLF}

From our binned luminosity function we derive the space density for $M_{1450}< -26.1$ (equivalent to the limit of $M_i < -27.6$ in \citealt{Richards:2006ye}) as: 

\begin{align*}
\rho(z = 3.8,M_{1450} < -26.1)_{\rm RL} = 3.57 \pm 0.44  \, {\rm Gpc}^{-3}\\
\rho(z = 4.2,M_{1450} < -26.1)_{\rm RL} = 1.54 \pm 0.63  \, {\rm Gpc}^{-3}
\end{align*}  
In \cite{Vigotti:2003vv} the RLF is assumed almost constant and is estimated as:
\begin{align*}
{\rm RLF}(M_{1450} < -26.9) = 13.3 \%  
\end{align*}

Therefore, for the total QSO population we obtain:

\begin{align*}
\rho(z = 3.8,M_{1450} < -26.1) = 26.8 \pm 3.3 \, {\rm Gpc}^{-3}  \\
\rho(z = 4.2,M_{1450} < -26.1) = 11.6 \pm 4.7 \, {\rm Gpc}^{-3} 
\end{align*}
in good agreement with the space densities derived from Richards et al. (2006): 

\begin{align*}
\rho(z = 3.75,M_{1450} < -26.1) = 29.0 \pm 2.0 \, {\rm Gpc}^{-3}  \\
\rho(z = 4.2,M_{1450} < -26.1) = 13.6 \pm 1.3 \, {\rm Gpc}^{-3} 
\end{align*}

\subsubsection{For redshift- and luminosity-dependent RLF}

\cite{Jiang:2007mi} use a sample of more than 30,000 optically selected QSOs from the SDSS  to study the evolution of the RLF as a function of redshift and luminosity.
 They find that the RLF of QSOs decreases with increasing redshift and decreasing luminosity, according to:
 
\begin{equation}
{\rm log} \frac{RLF}{(1-RLF)} = b_o + b_z log(1+z) + b_M (M_{2500} + 26)
\end{equation}

where $M_{2500}$ is the absolute magnitude at rest frame $2500$ \AA. 
The parameters $b_o$, $b_z$ and $b_M$ depend on the value of radio loudness and for $R>10$ ; they are $b_o = -0.132 \pm 0.116$, $b_z =  -2.052 \pm 0.261$, $b_M = -0.183 \pm 0.025$.  $M_{2500}$ is calculated from  $M_r$ as  

\begin{equation}
M_{2500} = M_r +2.5 \alpha_\nu \, {\rm log} \left (  \frac{2500 \, {\rm \AA}}{ 6231 \, {\rm \AA}} \right )
\end{equation}

We used the  \cite{Jiang:2007mi} formula to obtain  for each bin the corresponding value of the RLF  (column (5) in Table 5). The RLF lies in the range 3\%-10\%, and increases with decreasing $\phi_{\rm RL}$.  
Applying the corresponding RLF to each bin and integrating we obtain:

\begin{align*}
\rho(z \approx 3.8, M_{1450} < -26.1) = 81.7 \, \pm 31.7 ~ {\rm Gpc}^{-3}  \\
\rho(z \approx 4.2,M_{1450} < -26.1) = 41.0  \, \pm 31.1 ~ {\rm Gpc}^{-3} 
\end{align*}

This is a factor $\sim 3$ higher than the results from
\cite{Richards:2006ye}, but still within $2\sigma$, due to the large
errors in the luminosity function and in the RLF.  In particular, the
errors on the RLF at this redshift and magnitude are $\sim$
50\%. Given that our determination of the radio-loud luminosity
function agrees reasonably well with \cite{McGreer:2009nx} and with
\cite{Vigotti:2003vv}, this discrepancy cannot be attributed solely to
a possible overestimation of our luminosity function but may also be
due to to a systematic underestimation of the RLF in
\cite{Jiang:2007mi}. The large quoted errors invite caution when using the
\cite{Jiang:2007mi} formula to determine the fraction of radio-loud
quasars at high redshifts.

\subsection{The bright-end slope of the luminosity function for RL QSOs}

The QLF is usually well fitted by a double-power-law parametrisation that takes into account the redshift (e.g. \citealt{Pei1995}; \citealt{Peterson1997}; \citealt{Boyle:2000hc}; \citealt{Croom:2004ij}; \citealt{Richards:2006ye}):

\begin{equation}
\phi (L,z) = \frac{\phi^* / L^*}{ (L/L^*(z))^{-\alpha} + (L/L^*(z))^{-\beta}}
\label{doublePower}
\end{equation}
where $\alpha$, $\beta$, $\phi^*$ ,  $L^*$	are the faint-end slope, the bright-end slope, the normalisation of the luminosity function, and the characteristic break luminosity, respectively. This model, if $\alpha > \beta$, can be approximated by

\begin{equation}
\phi \propto
\begin{cases}
L^{\beta} & \text{if } L\gg L^*\\
L^{\alpha} & \text{if } L\ll L^*   
\end{cases}
\label{luminosityApprox}
\end{equation}

As already mentioned we calculated the LF in terms of optical
luminosity in two bins of redshift. We compare our results and the
best-fit slope with those of \cite{Richards:2006ye}
for the entire population of QSOs, and with the results of \cite{McGreer:2009nx} for RL
QSO with $R>70$.  The limiting magnitude of the QSOs samples used by
these authors ($M_{1450} < -26.1$ for \cite{Richards:2006ye} and
\cite{McGreer:2009nx}) was considered bright enough and far from the
break luminosity to approximate the LF by a single power law $ \propto
L^{\beta}$.

 This kind of approximation led in recent years to a long debate about an  apparent flattening of the bright-end slope for $z>4$, after it was noticed in early high-redshift surveys (\citealt{Schmidt1995}; \citealt{Fan2001}). These authors showed that the slope at $z> 4$ had a value $\beta \approx -2.5$, much shallower that the one seen at $z<2.2$  ($\beta = -3.3$, Croom et al. 2004). This flattening was then confirmed by  \cite{Richards:2006ye} who used a large, homogeneous QSO sample from the SDSS-DR3 extending to $z = 5$. At higher redshift the constraints are weaker as they come from small samples, but in general they do not confirm a continued flattening of the slope with increasing redshift. In fact, \cite{Willott2010}, combining the CFHQS (Canada-France High-z quasar survey) with the more luminous SDSS sample, derived the QLF from a sample of 40 QSOs at redshifts $5.74 < z < 6.42$ and found $-3.8 < \beta < -2.3$. At redshift $z \sim 6$, \cite{Jiang2008} find $\beta = -3.1 \pm 0.4$ using QSOs from SDSS Stripe 82.

Evolution of the shape of the QLF with redshift (changes in the slopes or in the location of the break luminosity) provides one of the fundamental observational constraints to the growth of super-massive black holes (SMBHs) over cosmic time.  Assuming that  brighter AGN have more-massive black holes, the flattening of the bright-end  would be a remarkable indication of a downsizing of the SMBHs at high redshift. Downsizing was reported also by X-ray surveys (\citealt{Ueda2003}; \citealt{Hasinger2005}; see also \citealt{Brusa2009}).

On the other hand, recent work by \cite{Shen2012} and
\cite{McGreer2013} aims to fill the gap in the QLF between $z\sim 3.5$
and $z\sim 6$, with the purpose of testing the flattening of the
bright-end slope at $z > 3$.  \cite{Shen2012} constrain the luminosity
function by Bayesian modeling and using an homogeneous sample of
SDSS-DR7 QSOs at $z = 0.3-5$. The results of \cite{Shen2012} and \cite{Richards:2006ye} 
are, in general, in good agreement, finding that the
curvature of the LF changes significantly beyond $z = 3$. However,
\cite{Shen2012} suggest that the apparent flattening of the slope
appears to be more related to a strong evolution of the break
luminosity than a change in the bright-end slope. A similar conclusion
is drawn by \cite{McGreer2013}, who find no evidence for
an evolution in the bright-end slope at $M_{1450} <-26$ for a 
sample of QSOs with $4.7 \le z \le 5.1$.  On the other hand,
\cite{McGreer2013} find evidence of strong evolution in the break
luminosity, as it brightens from $M^*_{1450}\approx -25.4$ at $z=2.5$
to $M^*_{1450} \approx -27.2$ at $z = 5$. They conclude that this
evolution could flatten the bright-end slopes for surveys where the
faint limit is near the break luminosity. \cite{McGreer2013} compared
different models for the evolution of the QLF normalization and break
luminosity. Eventually they found a good fit of their data with recent 
results from the
literature, using a modified version of a luminosity evolution and
density evolution (LEDE) model proposed by \cite{Ross:2012pi}. In
particular the evolution of the break luminosity in this model is
log-linear (up to $z \sim 5$), with a break luminosity that
brightens with redshift . This modified LEDE model predicts that for
$z \sim 3.8$ the break luminosity would be $M_{1450}^* \sim -26.2$ and
for $z \sim 4.2$ it would be $M_{1450}^* \sim -26.4$.

\begin{figure}
\centering
\includegraphics[width=0.48  \textwidth]{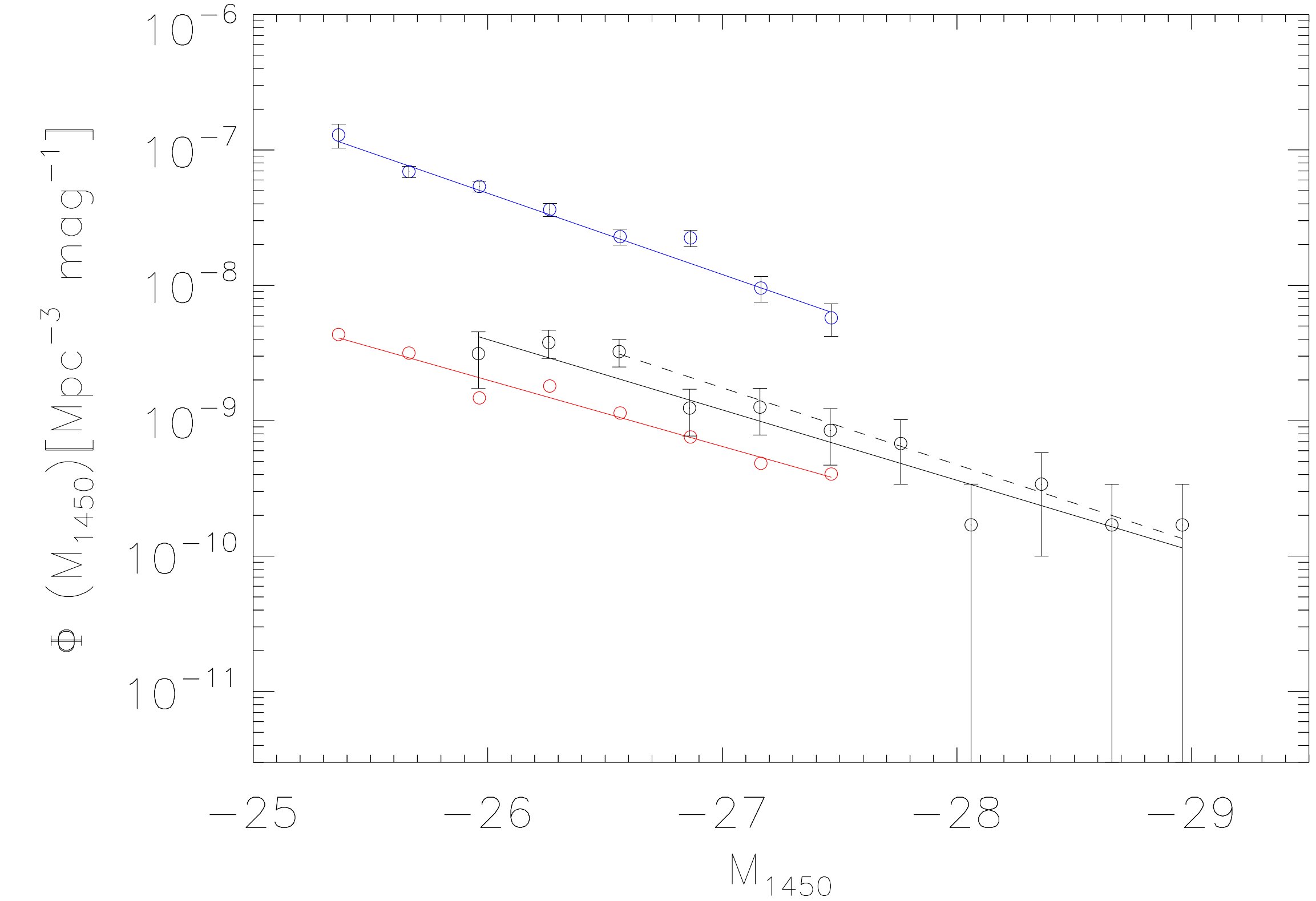}
\caption {Black points show the luminosity function derived for  $z \sim 3.8$, for RL QSOs with $log_{10} P_{1.4, \rm GHz} (W/Hz) > 25.7$  . For comparison, red points  show the luminosity function calculated by  McGreer et al. (2009), for RL QSOs with radio-loudness $R>70$, and  blue points the LF as calculated by Richards et al.(2006)  for the entire population of QSOs. Best-fit slopes are $\beta = -2.3$ for our LF (black line) ;  $\beta =-2.2$ for McGreer et al. 2009 (red line),  $\beta = -2.4$ for Richards et al.(2006) (blue line). For our LF we obtain $\beta = -2.3$ (black line) and $\beta = -2.4$ after excluding the two fainter points (dashed line). See Section 8.3.}
\label{slopes1}
\end{figure}

\begin{figure}
\centering
\includegraphics[width=0.48  \textwidth]{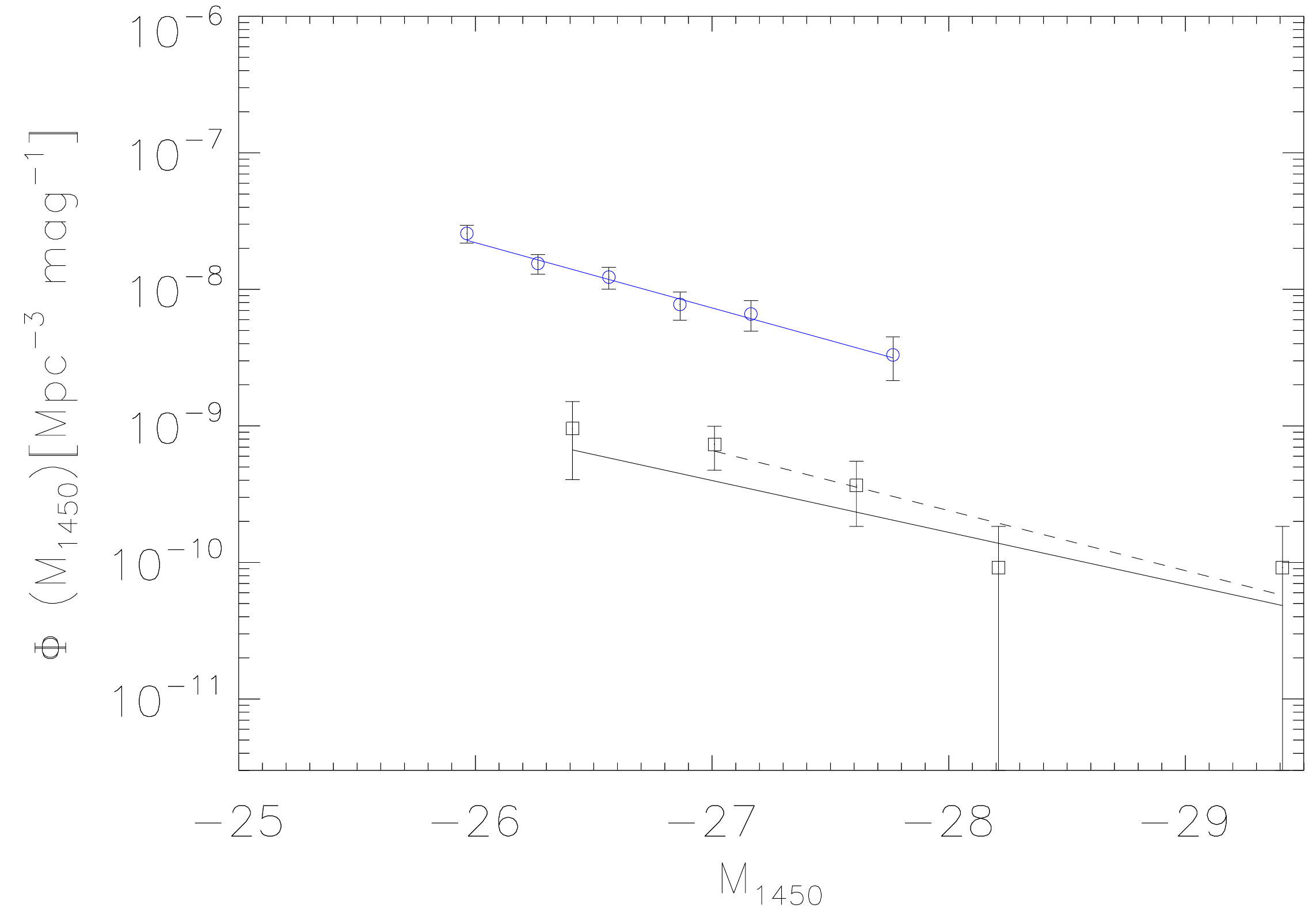}
\caption{Black points show the luminosity function derived for  $z \sim 4.2$, for RL QSO with $log_{10} P_{1.4, \rm GHz} {\rm (W/Hz)} > 25.7$  . For comparison, blue points show the LF as calculated by Richards et al. (2006) for the entire population of QSOs in the same bin of redshift. Best-fit slope is $\beta = -2.2$ for Richards et al. (2006) (blue line). For our LF we obtain $\beta = -2.0$ (black line) and $\beta = -2.1$ after excluding the faintest point (dashed line) See section 8.3}
\label{slopes2}
\end{figure}

  If we approximate the LF by a single power law $ \propto L^{\beta}$,
  we find that in the first bin of redshift, $z \sim 3.8$, our
  best-fit slope is $\beta = -2.3 \pm 0.2$. As shown in
  Fig.\ref{slopes1} our best-fit is in good agreement with the slope
  found in \cite{Richards:2006ye} ($\beta = -2.4 \pm0.1$), and
  with that found by \cite{McGreer:2009nx} ($ \beta = -2.2 \pm
  0.2$). For the RL-QLF calculated in our second bin of redshift,
  i.e. $z \sim 4.2$, we re-binned the LF using $\Delta M =0.6$, in
  order to reduce the statistical noise. In this way, as shown in
  Fig.\ref{slopes2} the best-fit slope is $\beta = -2.0 \pm 0.4$. This
  result is consistent with the result found by \cite{Richards:2006ye}
  for the entire population of QSOs, i.e.  $\beta = -2.2 \pm 0.1$.

Our determinations of the bright-end slope for the RL population of
QSOs at $z \sim 3.8$ and $z \sim 4.2$ are consistent with the
flattening (between these redshifts-bins) of the bright-end slope
found in \cite{Richards:2006ye} for $z \ge 4$, which
\cite{McGreer2013} suggest is due to a bias resulting from a
single-fit power law in a region near the break luminosity.

As we have quasars with luminosities near or below the predicted break
luminosity, we repeat the fit but excluding those points. In the first
bin at redshift $z \sim 3.8$, we exclude the two fainter points.  In
this way we obtain a slightly steeper best-fit slope and a larger
error: $\beta = -2.4 \pm 0.3$. This fit is shown in Fig.\ref{slopes1}
as a dashed line.  In the second bin of redshift, $z \sim 4.2$, we
exclude the faintest point, obtaining again a small increase of the
slope: $\beta = -2.1 \pm 0.4$. This fit is shown again as a dashed
line in Fig.\ref{slopes2}.

In light of the results from \cite{McGreer2013}, our data do not
strongly constrain the slope of the bright end nor the exact
location of the break luminosity, especially considering the large
errors of the brighter bins of the LF. Nevertheless this simple
derivation is consistent with the results of \cite{McGreer2013}.

In summary, our results are in good agreement with those of
\cite{Richards:2006ye} and \cite{McGreer:2009nx}. This result in
itself is not trivial, because we are comparing different populations
of QSOs in this range of redshift. In particular, we are comparing our
results with the whole population of QSOs (by comparing with
\citealt{Richards:2006ye}) and with a population of RL QSOs where the
radio-loudness is defined differently (being $R^*>70$ for the RL
sample of \citealt{McGreer:2009nx}). We therefore have indications of
a certain homogeneity of the QLF regardless of the differences in
radio-loudness.

On the other hand, since we can't constrain the bright and the
faint-end slopes, and we don't have an estimate of the break
luminosity, we can't conclude that the noted consistency of the
slopes imply also consistency of all the parameters. Any differences
of the slopes, or a different value of the break luminosity, could
point to different density evolution of the RL and the RQ populations.
In fact, \cite{Jiang:2007mi} express the dependence of the RLF on
optical luminosity as $\approx L^{0.5}$, implying $\beta _{RL
  QSO} \approx \beta_{QSO} + 0.5$, which is consistent with the
differences of slopes that we find between the first ($z\sim 3.8$) and
the second bin ($z\sim 4.2$) of redshift.  Also, \cite{Balokovi2012}
and \cite{kratzer2014}) find evidence that, at high redshift, the
radio-loudness distribution of quasars is not a universal function,
and likely depends on redshift and/or optical luminosity. We
therefore need a sample of RL QSOs at fainter luminosities to
constrain the faint-slope, and a larger survey area to extend the
bright end of the luminosity function and thus, determine the break
luminosity.

\section{Conclusions}

We construct a sample of high-redshift radio-loud QSOs $ 3.6 \le z \le 4$, and use it to measure the luminosity function and space density of QSOs in this range of redshift. Our principal conclusions are:

\begin{enumerate}

\item  We show (Section 3) that a simple neural network can be used to select high-redshift QSOs from radio-optical surveys, with $97$\% completeness and $60$\% efficiency.

\item With the aid of the neural-network, we construct a sample of 87
  radio-loud QSOs at redshift $\sim 4$. Of the various sources of
  incompleteness in the optical and radio surveys (Section 6),
  exclusion of SDSS 'CHILD' images is the main cause of our
  incompleteness relative to the SDSS selection of QSO candidates. But
  when applied to non-CHILD objects, our neural-network algorithm
  detects $\sim 97\%$ of the high-z QSOs, while SDSS only detect $\sim
  85\%$ of them.

\item We determine the optical luminosity function for radio-loud QSOs in two redshift bins, $3.6 \le z < 4.0$ and $4.0 \le  z \le 4.4$ (Section 7), and measure the total
comoving density of QSOs in these two redshift ranges (Fig. 12), obtaining a result consistent with that of \citet{Vigotti:2003vv} at luminosities $M_{\rm 1450}<-27.6$. We also find good agreement between our cumulative luminosity functions (Fig. 13) and that measured by \citet{Vigotti:2003vv} and \citet{Carballo:2006uq}, which determine the space density at intermediate redshifts.

\item Assuming a radio-loud fraction of $13.3$\%
  \citep{Vigotti:2003vv} we estimate the total comoving density of
  QSOs (Section 8.2.1). The derived density of QSOs at $z\sim 4$ is
  consistent with that of \cite{Richards:2006ye}.  Alternatively
  (Section 8.2.2), using the redshift- and luminosity-dependent
  radio-loud fraction found by \cite{Jiang:2007mi}, we measure a total
  comoving density of QSOs a factor 3 higher than measured by
  \cite{Richards:2006ye}. However, this result is significantly affected by
  the large error bars on the formula assumed for the radio-loud fraction (RLF).

\item We determine the slope of the luminosity function in two bins of redshift (Section 8.3). In the lower-redshift bin (z = 3.8)
we found $\beta = -2.3 \pm 0.2$, consistent with \cite{Richards:2006ye} and \cite{McGreer:2009nx}. In the higher redshift bin ($z = 4.2$) we find a slope $\beta -2.0 \pm 0.4$ consistent with \cite{Richards:2006ye}. Values of the slope consistent with our determination, have been interpreted as a flattening of the bright end slope for the high-z QSOs population, but has recently been re-interpreted as the result of a strong evolution of the break luminosity for high-z QSO \cite{McGreer:2009nx}.The consistency of our results with \cite{Richards:2006ye} and \cite{McGreer:2009nx} suggests a similar evolution for both radio-loud and radio-quiet populations. Our results can be also interpreted as suggestive of a flattening of the bright-end slope from $z\sim 2$ to $z\sim 4$, for the radio-loud population only. If confirmed, this implies an evolution of the density of super-massive black holes associated with radio-loud QSOs, in the sense that they were more abundant at $z \sim 4$. However, to clarify the evolution of the RL population relative to that of the whole population of QSOs, more observational constraints are needed, especially at redshifts above 4. The candidate-selection approach described here is now being applied to FIRST-SDSS-UKIDSS surveys to search for QSOs at $z\gtrsim 4.5$ (Tuccillo, McMahon \& Gonz\'alez-Serrano, in prep.)

\end{enumerate}

\section*{Acknowledgements}

This work has been funded by the Spanish Ministerio de Ciencia e Innovaci\'on (MICINN) under project
AYA2011-29517-C03-02. This article is based on observations made with the Nordic Optical Telescope operated
 by the Nordic Optical Telescope Scientific Association in the Spanish Observatorio del Roque de 
los Muchachos on the island of La Palma. Funding for the creation and distribution of the SDSS Archive has been provided by the Alfred P. Sloan Foundation, the Participating Institutions, the National Aeronautics and Space Administration, the National Science Foundation, the U.S. Department of Energy, the Japanese Monbukagakusho, and the Max Planck Society. The SDSS Web site is http://www.sdss.org/.
The Participating Institutions are The University of Chicago, Fermilab, the Institute for Advanced Study, the Japan Participation Group, The Johns Hopkins University, the Max-Planck-Institute for Astronomy (MPIA), the Max-Planck-Institute for Astrophysics (MPA), New Mexico State University, Princeton University, the United States Naval Observatory, and the University of Washington. This research has made use of the NED which is operated by the Jet Propulsion Laboratory, California Institute of Technology, under contract with the National Aeronautics and Space Administration. We thank R. Carballo for the Neural Network software and for helping to train the network  and validate the selection of QSOs candidates.
We acknowledge the helpful discussions with Richard G. McMahon during early stages of this work. Finally, we thank our anonymous referee for giving constructive comments, which substantially helped improving the article.

\bibliography{mybib}
\bibliographystyle{mn2e}
\end{document}